\documentclass[twocolumn, superscriptaddress, amsmath, amssymb, aps]{revtex4-2}

\usepackage{graphicx}
\usepackage{dcolumn}
\usepackage{bm}
\usepackage{xcolor}
\usepackage{mathtools}
\usepackage{physics}
\usepackage[normalem]{ulem}
\usepackage{soul}
\usepackage{braket}

\usepackage{hyperref} 
\hypersetup{colorlinks=true, citecolor=blue}

\begin{document}


\title{Asymmetric Transmission Through a Classical Analogue of the Aharonov-Bohm Ring}

\author{Lei Chen}
 \affiliation{Maryland Quantum Materials Center, Department of Physics, University of Maryland, College Park, Maryland 20742, USA}
 \affiliation{Department of Electrical and Computer Engineering, University of Maryland, College Park, Maryland 20742, USA}

\author{Isabella L. Giovannelli}
 \affiliation{Maryland Quantum Materials Center, Department of Physics, University of Maryland, College Park, Maryland 20742, USA}

\author{Nadav Shaibe}
 \affiliation{Maryland Quantum Materials Center, Department of Physics, University of Maryland, College Park, Maryland 20742, USA}
 
\author{Steven M. Anlage}
 \email{Corresponding author: anlage@umd.edu}
 \affiliation{Maryland Quantum Materials Center, Department of Physics, University of Maryland, College Park, Maryland 20742, USA}
 \affiliation{Department of Electrical and Computer Engineering, University of Maryland, College Park, Maryland 20742, USA}

\date{\today}

\begin{abstract}


It has been predicted that new physics and technology are enabled for quantum systems that suffer from partial decoherence, in the intermediate range between coherent quantum evolution and incoherent classical physics.  We explore the asymmetric transmission through a classical  analogue of the Aharonov-Bohm (AB) mesoscopic ring that supports a 3:1 asymmetry in transmission times, augmented with lossy features that act preferentially on the longer-lingering  waves.  Such a device is realized as a linear microwave graph utilizing a gyrator to create the 3:1 transmission time delay asymmetry, along with both homogeneous and localized losses, to produce an imbalance in wave transmission through the device.  We demonstrate asymmetric transmission through the microwave-ring graph as a function of loss in both simulation and experiment, and in both the frequency- and time-domain.  The microwave ring-graph results are compared to a numerical simulation representative of a class of recent models proposing dephasing-induced transport asymmetry in few-channel quantum systems, and parallels are noted.\\


\end{abstract}

\maketitle









\section{Introduction}
There is a deep disparity between the quantum and classical perspectives on physical laws.  Our experience with the classical world is dominated by equations of motion that include loss, dissipation, and friction, explicitly breaking time-reversal invariance, and describing phenomena involving large numbers of interacting particles, creating a perception of the ``arrow of time" \cite{Jar11,Seifert12}.  At the microscopic scale we know that the laws of quantum mechanics are time-reversal invariant, and isolated systems are described by  unitary time evolution of the quantum state.  This poses the question: how do the laws of quantum mechanics segue into those of classical mechanics as one admits stronger interactions with other degrees of freedom?  Such questions have been addressed in the fields of fundamental quantum mechanics \cite{Zurek91,Poya96,Joos03,Zurek03,Fort19,Braak20}, quantum chaos \cite{Casati21}, quantum statistical mechanics \cite{Jar11} and thermodynamics \cite{Breuer02}, and mesoscopic physics \cite{Ast02,Mann21}, among others.  

A related question is whether or not there is new physics to explore in the regime between pure quantum evolution, and classical physics?  Can systems described by a mixture of quantum and classical properties show qualitatively new phenomena that are not anticipated by the properties of systems in either limit?  More specifically, can a finite degree of quantum `dephasing' be harnessed to perform a new task that is not possible in the fully quantum or fully classical limits?  The answers to these questions are relevant to a wide variety of phenomena and applications.  For example, coupling a quantum system to additional degrees of freedom can be used for quantum state preparation,\cite{Kraus08} as a resource for universal quantum computation,\cite{Vers09} to simulate open quantum systems,\cite{Barr11,Mull12,GPIBMQ20}  to enable environment-assisted quantum transport (ENAQT),\cite{Plenio08,Reb09,Caruso09,Maier19,Zerah20} to accomplish fast reset of qubits,\cite{Val06} and for advanced quantum sensing \cite{Cabot24}.  However, it is interesting to see if other qualitatively new phenomena can be discovered and exploited in this regime.

There has been long-sustained interest in preparing materials and devices that explicitly break time-reversal invariance (TRI) to create non-reciprocal transport for electrons between equivalent contacts \cite{Ast02,Mann21}, for example.  Many proposals to create non-reciprocal transport are inspired by Maxwell demons,\cite{Max72,2ndLaw05} thermally-driven ratchets,\cite{Reim02}  non-centro-symmetric materials,\cite{Tokura18} and efforts to violate detailed balance in thermal radiation \cite{FanDetBal,FanDetBal2}.  Systems with nonlinearity have been utilized to demonstrate asymmetric transport \cite{Casati11,Bend13}.  One approach to selectively transferring electrons is to create wavefunction interferometers that preferentially pass matter waves moving in one direction but not the other, and such devices require breaking of time-reversal invariance \cite{Mann18,BredSol21}.  In particular, the Aharonov-Bohm matter-wave interferometer, augmented with decoherence to induce non-reciprocity,\cite{Bred21} has been proposed as one possible setting for this type of device.

The Aharonov--Bohm (AB) effect is a quantum mechanical phenomenon in which an electrically charged particle of charge $q$ is affected by an electromagnetic potential (through the vector potential $\vec{A}$) in the absence of a magnetic or electric field at the location of the particle \cite{AharonovBohm1959}. This effect gives rise to a type of ``action at a distance'' in which a particle can be affected by electromagnetic fields even if it does not experience them directly.  The traditional setting for the AB effect is a finite region of space containing non-zero magnetic flux $\Phi$ surrounded by a field-free region.  A beam of charged particles is sent through the field-free region in a ring geometry that fully incorporates the region of finite flux within the ring.  The charges show interferometric properties arising from a quantum phase shift $e^{\frac{i q}{\hbar} \int \vec{A}\cdot d\vec{\ell}}$ upon traversing each branch of the ring.  Upon recombination, parts of the wave packet that pass through different branches of the ring create interference that depends on $\Phi/\Phi_0$, where $\Phi_0=h/q$ is the flux quantum for the charges, and $h$ is Planck's constant.

\begin{figure}[ht]
\includegraphics[width=0.48\textwidth]{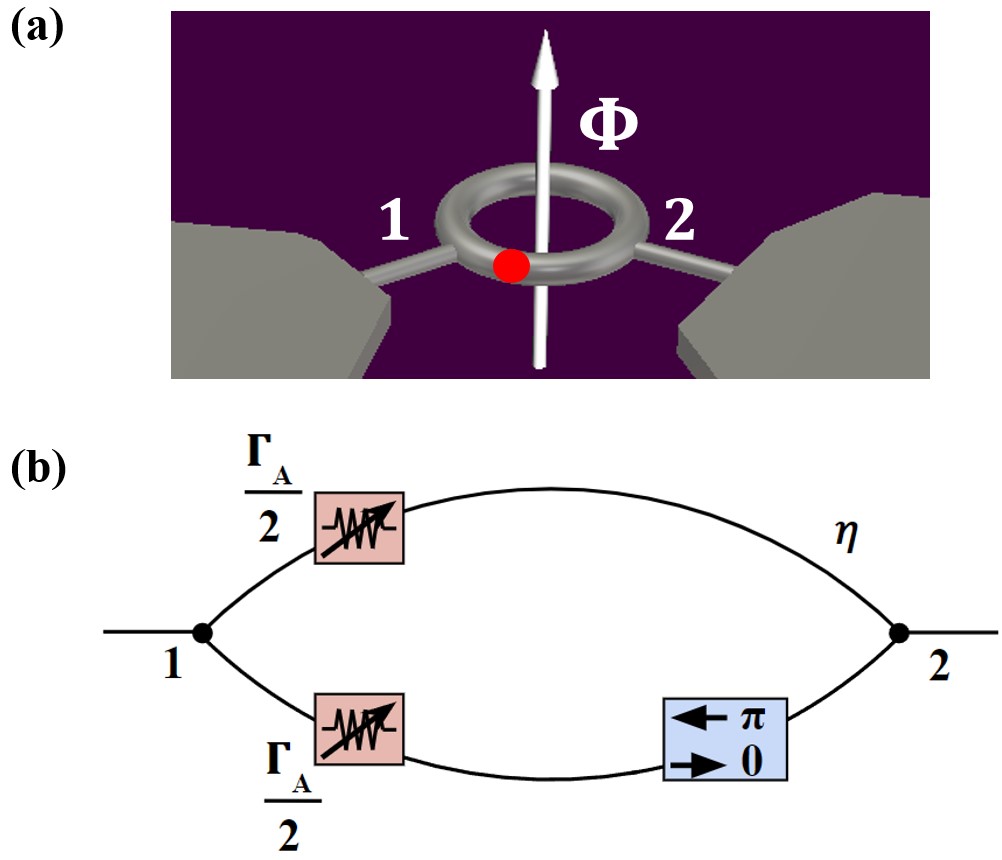}
\caption{Schematic layout of an Aharonov--Bohm ring and a corresponding classical microwave graph analogue. (a) shows the device layout of a two-port Aharonov--Bohm ring made of clean metal. The penetrating magnetic flux $\Phi$ creates non-reciprocal transmission times for electron wave packets transiting the ring. The red dot marks an inelastic scattering center. (b) shows the classical analogue microwave graph realization. A microwave gyrator (see Supp. Mat. Fig. \ref{gyrator}) creates the $\pi$ phase difference for waves travelling in opposite directions on the lower bond. The variable attenuators of strength $\Gamma_A/2$ act as a localized loss center.  Panel (a) after Ref. \cite{Man2020}.  }
\label{schematic_ABRing}
\end{figure}

In Refs. \cite{Mann18,ManBr19,BredSol21,Man2020}, the authors consider an 
Aharonov-Bohm (AB) ring in a mesoscopic conducting sample that suffers from a certain degree of ``dephasing" in electron transport. Specifically, they consider a single-channel, 2-terminal AB ring enclosing a DC magnetic flux (as described above). 
Fig. \ref{schematic_ABRing}(a) shows the schematic of the proposed mesoscopic Aharonov--Bohm ring.  This device has two key features that are proposed to bring about new behavior. First, this device has non-reciprocal transmission times as a result of the direction dependent phase shift that the electron picks up as it traverses the ring. An electron wavepacket propagating through the device from left to right, say, will split into two parts at the left combiner. Along each of these paths, the respective wave will accumulate a phase. In order for the waves to constructively recombine at the other end, they must be in phase with each other, hence the net phase difference accumulated must be $2\pi n, n\in\mathbb{Z}$. The phase difference accumulated as the electron traverses the ring is given by,\cite{BredSol21} $\Delta\phi=k \Delta l \pm 2\pi\frac{\Phi}{\Phi_0}$
where $\Delta l$ is the difference in length between the two arms and $k$ is the wavenumber of the electron. 
In \cite{BredSol21}, the author considers the case where $k\Delta l = \frac{\pi}{2}$ and $2\pi\frac{\Phi}{\Phi_0}=\frac{\pi}{2}$, leading to $\Delta\phi_{1\rightarrow 2} = 0$ and $\Delta\phi_{2\rightarrow 1} = \pi$. Thus, electrons traveling from port 1 to port 2 would suffer no relative phase shift and immediately exit the ring. However, electrons traveling from port 2 to port 1 will suffer destructive interference, due to the $\pi$ phase difference, causing the wave packet to reflect back around the ring. Hence an electron starting at port 2 will need to traverse the ring 3 times in order to accumulate a net phase difference of $2\pi n, n\in\mathbb{Z}$, allowing it to coherently recombine and exit the ring. This consequently leads to a non-reciprocal 3 to 1 ratio of transmission times, as further discussed in \cite{Man2020}.

The second key feature arises when a finite degree of quantum dephasing is added in the form of a defect or ``deep trap" in the semiconducting ring \cite{ManBr19,Mann21,Bred21}, denoted as the red dot in Fig. \ref{schematic_ABRing}(a). This defect works as a ``dephasing site" by taking a passing electron wavepacket out of the conduction band with some probability. When an electron is captured, the state is projected onto an eigenstate of the trap and it loses all memory of its prior state, including its phase and the direction that it was traveling. The electron is eventually released back into the conduction band with equal probability of traveling to the left or right. A more detailed description of this process is given in \cite{Mann21}. It is argued in \cite{Mann21, Man2020} that this dephasing site, combined with the engineered 3 to 1 transmission time ratio, leads to a net asymmetry in transmission probability. The argument is that since electrons traveling from port 2 to port 1 need to make three times as many passes as electrons traveling from port 1 to port 2, they then have three times as many chances of interacting with the dephasing site. Hence electrons traveling in that direction have a lower chance of being transmitted.  
Supporting simulation results are discussed in \cite{Mann21} where they show that not only is this asymmetry in transmission probability present, but it has a non-monotonic dependence on the modelled dephasing rate.


In addition to the explicit proposals made in Refs. \cite{Mann18,ManBr19,Man2020}, a few other works have hinted at this asymmetric transmission probability effect.  The work of Entin-Wohlman, \textit{et al.} essentially considers a quantum ring subjected to an external magnetic field and suffering partial dephasing in electron transport around the graph.  They calculate that this results in a loss of detailed balance in equilibrium, and the creation of a net current between two points in the ring \cite{EW95}.  A cold atom version of a \textit{dissipative} AB-ring in momentum space, roughly similar to the quantum models cited above \cite{ManBr19,Mann21,Bred21}, has demonstrated nonreciprocal quantum transport \cite{ColdAtom20}.  This approach utilizes a synthetic magnetic flux and laser-induced \textit{loss of atoms} in a Bose-Einstein condensate to break inversion and time-reversal symmetries, and demonstrates directional atom flow.

In this paper, our objective is to examine the properties of a \textit{classical analogue} of the mesoscopic AB-ring proposed in \cite{Mann18,ManBr19,Man2020}. In particular, we are motivated by their proposition that quantum wave interference, combined with finite dephasing can lead to asymmetric transmission. Our goal is to see if parallel behavior can be observed in a purely classical setting. Specifically, we want to understand if classical wave interference effects, combined with dissipation, give rise to asymmetric wave propagation.

The outline of the paper is as follows.  We begin by discussing the background for the experiment, where we provide a brief overview of the parasitic channel phenomenology that provides an organizing principle for understanding the effects of loss on our experimental results.  We then present simulations of the microwave-ring graph and demonstrate the properties required to achieve asymmetric transmission.  A number of complications are added to the simulations to represent the experimental situation, and the key properties are shown to survive.  Next we present the experimental realization of the microwave ring-graph, and discuss measurements of the scattering matrix and time delays in both the frequency and time domains, and find good consistency with the simulations.  Finally we present results on the transmission asymmetry through the graph in experiment, and compare to expectations based on simulations.

\section{Background}
There is a long history of using microwave circuits and resonant microwave billiards to emulate the wave interference properties of quantum systems \cite{Stock99,Stock01}.  Quantum billiards are simulated using microwave resonators, where the analogy between the Schr\"{o}dinger equation for the wavefunction $\psi$ and the electromagnetic Helmholtz equation for one component of electric field $\vec{E}$ is particularly clear in two-dimensions \cite{McD79,Stock90,Doron90,Sridhar92,Sridhar95,Richter01,Bird1,Bird2,Drago04,Bird3}.  For example, it was also shown that there is an analogy between the Poynting vector for energy flow in the quasi-two-dimensional electromagnetic cavity and the probability current in the corresponding quantum system \cite{Seba99}.  Similarly, the analogy between solutions to the Schrodinger equation on multi-connected molecules \cite{Pauling36} and the wave equation on networks of microwave graphs has been noted \cite{Hul04}.

In this paper we will be adopting the Heidelberg picture of wave scattering \cite{Agassi75,VWZ85,Weid92,Fyodorov97,FyoSavSomRev05,MRW2010,FSav11,KuhlEHF13,Schomerus2015}  
to describe scattering in non-Hermitian systems. 
Localized loss is modeled by a matrix $\Gamma_A$ that evaluates the overlap of the $M$ resonant eigenmodes of the closed system at the location of loss, represented by a set of $L$ absorption channels. 
The $\Gamma_A$ matrix is added to the Hamiltonian of the closed system to create an effective Hamiltonian that is non-Hermitian.  In addition, one can model uniform attenuation $\eta$ by evaluating the energy (or equivalently frequency of the microwaves) with an imaginary offset $E \rightarrow E+i\eta$. We note that this is a simple phenomenological approach to describing loss that is quite generic, and therefore applies to all sorts of classical wave scattering systems (electromagnetic, acoustic, mechanical, etc.).  However, it is not a microscopic theory, and in the case of quantum systems suffering from decoherence, it altogether fails to capture the quantum degrees of freedom, and their associate quantum interactions with the scattering system.  As such, the Heidelberg picture has no microscopic justification in describing the scattering properties of quantum systems suffering from decoherence.

In Ref. \cite{Sameer06} it was experimentally demonstrated that the statistical properties of the scattering matrix of a microwave cavity with variable uniform loss were described by a simple dephasing-lead  model borrowed from a treatment that utilized the parasitic channel model of dephasing in mesoscopic transport \cite{Butt86,BB97,Ben02}.  The uniform attenuation of the quasi two-dimensional microwave cavity (parameterized as the dimensionless quantity $\alpha = k^2A/4\pi Q$, where $k=\omega/c$, $\omega$ is angular frequency, $A$ is the area of the billiard, and $Q$ is the typical quality factor of the resonant modes) was found to be directly proportional to the dimensionless phenomenological dephasing rate $\gamma$ that governs the statistics of the scattering (S) matrix and conductance in the corresponding mesoscopic billiard \cite{BB97,Ben02}, as $\gamma = 4 \pi \alpha$ over a wide range of microwave loss.  (It can be shown that $\alpha$ is related to the uniform attenuation as $\alpha = \frac{\eta}{2\Delta}$, where $\Delta$ is the mean spacing between modes of the closed system \cite{Sameer06,Chen21}.)  This demonstrated that the statistical fluctuations of the wave interference properties of both the classical microwave experiment, and the model quantum system, have the same dependence on a single parameter.  We refer to this single-parameter dependence of the scattering matrix statistical properties as the ``\textbf{parasitic channel phenomenology}" (\textbf{PCP}).  We shall explore the PCP further in this paper through a study of wave interference effects in a classical analogue of the AB-ring graph.  

Here we consider quasi-one-dimensional microwave network analogues of quantum graphs that emulate corresponding few-channel mesoscopic conducting quantum systems.  Quantum graphs are one-dimensional metric graphs with complex topology that support excitations described by the Schrodinger operator \cite{Kottos1997,Kottos99} that can be used to describe quantum transport through a variety of structures, including molecules \cite{Pauling36,Net72}, quantum wires \cite{Exner95,San98}, disordered two-dimensional quantum dots \cite{Chalk88,Kowal90,San98}, etc.  In general, microwave graphs consist of vertices connected by bonds, and we consider the propagation of microwave signals on the bonds and the resulting interference that occurs when multiple bonds meet at a vertex.   Microwave graphs have been used to investigate many aspects of wave scattering theory, including statistics of the Wigner reaction matrix (analogous to electromagnetic impedance) \cite{Hul04}, topological edge invariants \cite{HuTopo15}, deviations from the predictions of random matrix theory \cite{Dietz17}, etc.  The PCP has also be used to quantify the effects of dissipation on the scattering matrix statistics of microwave graphs \cite{Hul05,Ghut23}.

We introduce a linear two-terminal, single-channel microwave graph to emulate the wave interference properties of a mesoscopic Aharonov--Bohm ring, and study its scattering properties in both the frequency domain and the time domain. In particular, our objective is to demonstrate asymmetric transmission by utilizing a specific combination of wave interference and attenuation.  The microwave graph in Fig. \ref{schematic_ABRing}(b) satisfies all of the conditions for displaying asymmetric transmission, including broken time-reversal invariance, a sub-unitary scattering matrix, and a gyrator which produces a broadband constant $\pi-$phase shift for waves going in one direction \cite{ManBr19}.  We note that the gyrator effectively performs the function of a Faraday rotator, as originally utilized by Rayleigh to create non-reciprocal transmission of light \cite{Mann21,Strutt85,Rayleigh01}.  We demonstrate that the real part of forward and backward transmission time delays have a 3:1 ratio in a broad range of frequency, which establishes the first condition for asymmetric transmission. We also introduce a localized loss $\Gamma_A$ (exploiting the PCP to phenomenologically treat dissipation \cite{Sameer06}) in the ring graph, and adjust the left/right transmission coefficients by varying the attenuation $\Gamma_A/2$ of two objects, in a balanced manner. We demonstrate the dependence of asymmetric transmission on the loss rate through the attenuation variation.




\section{Simulations}
We first simulate the non-reciprocal transmission effect of the microwave ring-graph in CST Microwave Studio. In particular we use a circuit modeling package to create a faithful model of the microwave graph (see Appendix \ref{sec:CST} and Fig. \ref{schematic_Att}).  We implement the complete circuit shown in Fig. \ref{schematic_ABRing}(b) with equal electrical lengths (i.e. propagation phase shifts) for the upper and lower bonds.

\subsection{Frequency Domain Simulations}
\begin{figure}[ht]
\includegraphics[width=0.48\textwidth]{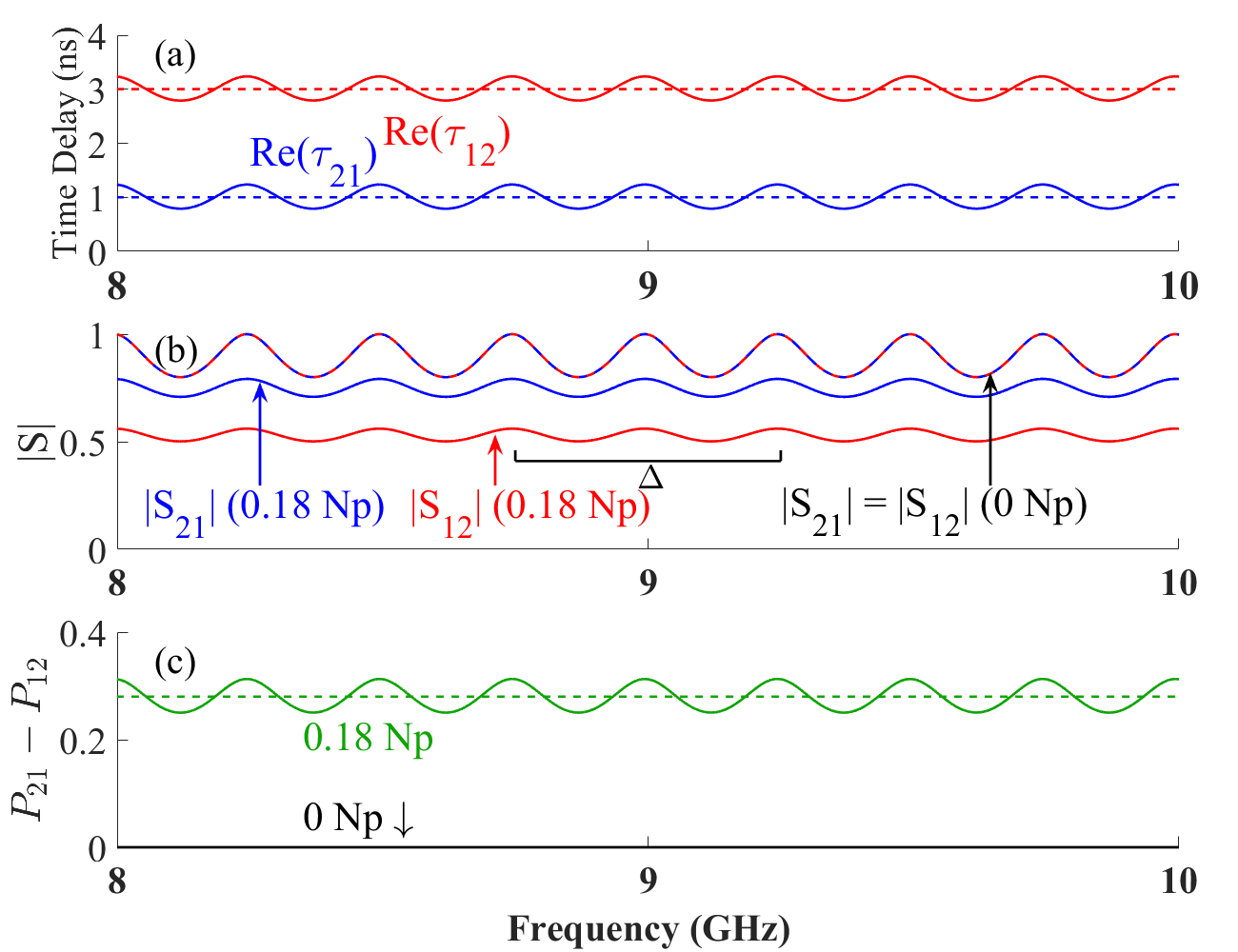}
\caption{Simulated properties of the microwave AB-ring graph with uniform attenuation $\eta=0$.  (a) Real part of transmission time delays of the Aharonov-Bohm microwave ring graph simulation for the case of localized attenuation $\Gamma_A/2 = 0.18$ Np. Blue curve shows $\text{Re}[\tau_{21}]$ and red curve shows $\text{Re}[\tau_{12}]$, and the dashed lines show the mean values, demonstrating a 3:1 ratio. (b) Linear transmission magnitudes in both directions for the case of $\Gamma_A/2 = 0$ Np attenuation and $\Gamma_A/2 = 0.18$ Np local attenuation. Blue corresponds to $|S_{21}|$ while red corresponds to $|S_{12}|$. With no attenuation the transmission magnitudes are identical. The scale bar gives the expected periodicity frequency scale ($\Delta = 0.5$ GHz) for the shape resonances of a simple ring graph. (c) The asymmetric transmission $P_{21}-P_{12}$ as a function of frequency for the case of $\Gamma_A/2 = 0.18$ Np local attenuation. }
\label{S_ABRingGraph_Att}
\end{figure}

The complex transmission time delays (see Appendix \ref{sec:CTD}) for both directions ($S_{21}$ and $S_{12}$) can be calculated from the complex scattering matrix data as $\tau_{T}^{12,21} =-i \frac{d}{df} \log(S_{12,21})$ \cite{Asano16,Chen2021gen,dH21,BredSol21,Genack21,Gen22,Chen2022}.  Figure \ref{S_ABRingGraph_Att}(a) shows the simulation comparison between the real part of the transmission time delay for the two directions, demonstrating the required 3:1 ratio, modulo small oscillations associated with the shape resonances of the microwave graph \cite{Chen2022}.  Fig. \ref{S_ABRingGraph_Att}(b) shows the transmission magnitudes $|S_{12}|$ and $|S_{21}|$ through the microwave graph under several different attenuation settings.  When the lumped attenuation $\Gamma_A/2$ is 0 Nepers and there are no uniform losses, i.e. no loss in the entire system and a unitary S-matrix, the two transmission paths have identical transmission magnitude as a function of frequency.  As the lumped attenuation increases, significant  differences begin to show up between the two transmission amplitudes. 

The asymmetric transmission behavior of the ring graph at each frequency can be quantified as the transmission probability difference: $P_{21}-P_{12} = |S_{21}|^2 - |S_{12}|^2$.
Note that $P_{21}-P_{12}$ is bounded between 0 and 1, and is frequency dependent in general.  If there is no asymmetric transmission then $P_{21}-P_{12}=0$, while a non-zero value implies some degree of asymmetric transmission.  Fig. \ref{S_ABRingGraph_Att}(c) shows $P_{21}-P_{12}$ as a function of frequency for the $\Gamma_A/2 = 0.18$ Np attenuation case, revealing a frequency-averaged value (dashed green line) of $\braket{P_{21}-P_{12}}=0.28$.


\begin{figure}[ht!]   
\includegraphics[width=0.49\textwidth]{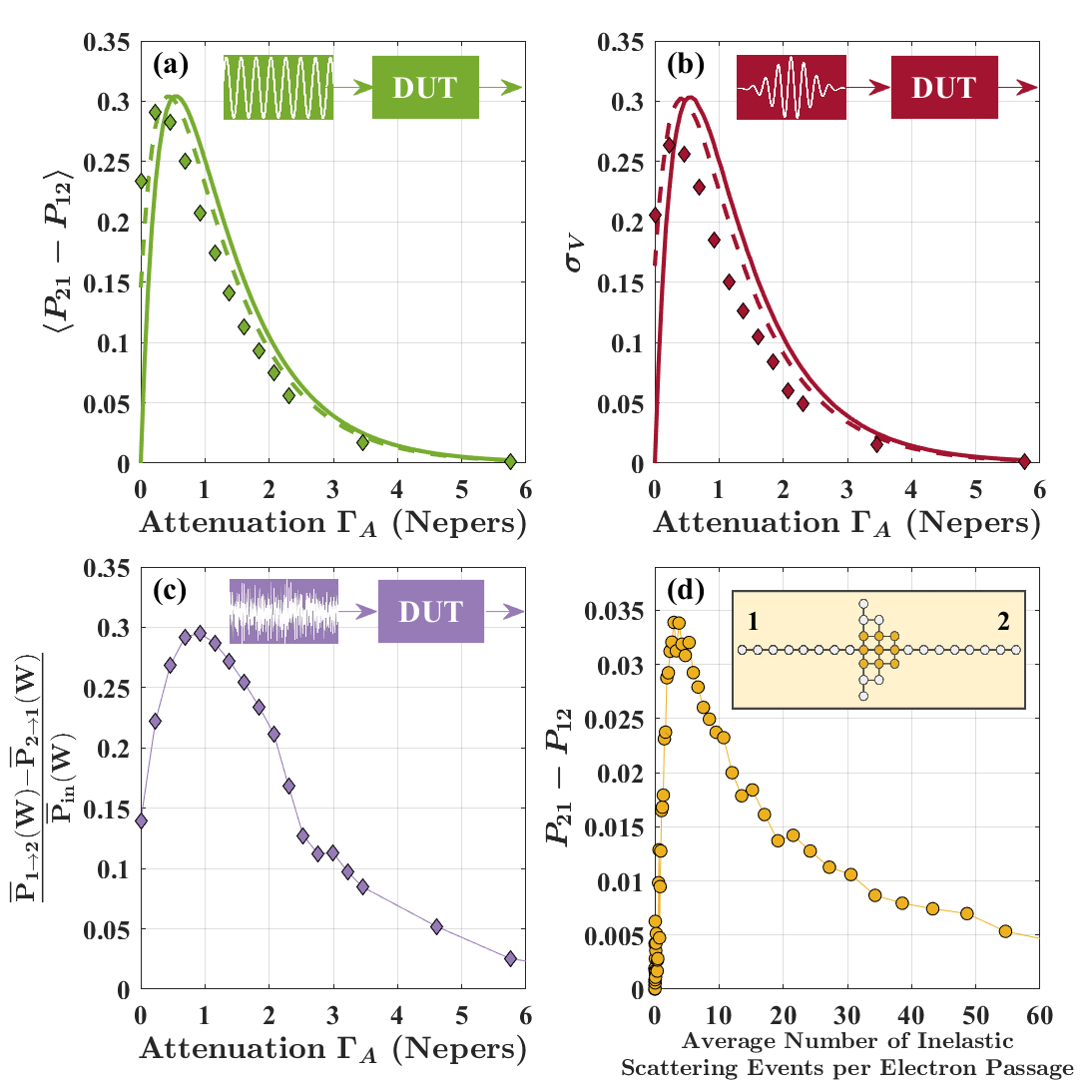}
\caption{Asymmetric transmission through microwave AB-ring graph and a model mesoscopic electron device.(a) Summary of the frequency domain results. In the upper right is a schematic of how these measurements work. For frequency domain measurements, a wave with a single frequency is sent through the device, this is done multiple times over a frequency range. For all curves, the quantity $P_{21}-P_{12}$ is averaged over the frequency range $8.25 - 8.75$ GHz, which is denoted as $\braket{P_{21}-P_{12}}$. The solid line is from simulation where only the localized attenuation from the attenuators is considered. The dashed line is also from the same simulation but with finite uniform attenuation added. The discrete diamond points are from experimental data. (b) Summary of the time domain results expressed in terms of $\sigma_V = (|V_{21}|^2 -|V_{12}|^2)/|V_{in}|^2$. For these time domain measurements, a 1 ns wide Gaussian pulse with a carrier frequency of 8.5 GHz is sent into the device. The legend has the same meaning as in (a). (c) Noise source input experimental results. This data was taken in the frequency domain with a microwave noise source input as a function of total attenuation in the graph. It is plotted as frequency-averaged transmitted power difference, normalized by the measured incident power, $(\bar{P}_{21}-\bar{P}_{12})/\bar{P}_{in}$. (d) Tight-binding model of triangle-shaped mesoscopic device with a dephasing center, as a function of the average number of inelastic scattering events per electron passage through the structure.  From Refs. \cite{Mann21,Bred21}.  Data provided courtesy of Dr. Jochen Mannhart, Max Planck Institute for Solid State Research, Stuttgart, Germany.}

\label{dephasing}
\end{figure}



To explore the asymmetric transmission further, Fig. \ref{dephasing}(a) shows $P_{21}-P_{12}$ as a function of total lumped attenuation $\Gamma_A$ for the model graph. Due to the periodic wiggles arising from the shape resonances of the ring graph, we perform an average of $P_{21}-P_{12}$ over a range of frequencies corresponding to one period of the shape resonances $\Delta = 0.5$ GHz, and designate it as $\braket{P_{21}-P_{12}}$.  The frequency averaged transmission asymmetry $\braket{P_{21}-P_{12}}$ shows a non-monotonic bell shaped behavior as the lumped attenuation increases.  

We have also simulated the more realistic case where both uniform attenuation and lumped variable attenuation are present in the AB-ring microwave graph.
We set the parameters of the simulated coaxial cables to approximately match those used in the experiments discussed below, leading to a frequency dependent uniform attenuation described by Eq.~\ref{attneqn}.  The resulting transmission asymmetry $\braket{P_{21}-P_{12}}$ vs. lumped attenuation is also shown in Fig. \ref{dephasing}(a) as a dashed line.  The main difference is that uniform attenuation effectively establishes a finite amount of asymmetric transmission even with zero lumped attenuation.

\subsection{Time Domain Simulations}
Time-domain simulations have been performed with the model microwave ring-graph shown in Fig. \ref{schematic_Att} (see Appendix \ref{sec:CST} and Supp. Mat. section ``Simulations"). 
 Gaussian wave packets with 1 ns width and a chosen carrier frequency are injected into the graph from each port.  The output pulses show the expected 3:1 asymmetry in time delay upon going through the graph, as shown in Fig. \ref{Sim_Pulse}.  To quantify the transmission probability difference in the time-domain, we form the quantity $\sigma_V = (|V_{21}|^2 -|V_{12}|^2)/|V_{in}|^2$, where $V_{21}$ and $V_{12}$ are the voltage amplitudes of the transmitted pulses in the simulations and $V_{in}$ is the amplitude of the incident pulse.  Fig. \ref{dephasing}(b) shows $\sigma_V$ as a function of the lumped attenuation (solid line).  As the lumped attenuation increases the transmitted wavepackets first show an increasingly asymmetric transmission probability.  However, beyond about $\Gamma_A=1/2$ Np the wavepackets show reduced transmission asymmetry.  The addition of uniform loss serves to shift the $\sigma_V$ curve (dashed line), just as with the frequency domain simulations in Fig. \ref{dephasing}(a).


These simulation results of the microwave ring-graph in both the frequency- and time-domains establish the basic properties of asymmetric transmission.

\section{Experiment}

The microwave graph consists of a coaxial cable structure that supports a single mode of propagation for microwave frequencies below the cutoff of higher-order modes, which is well beyond our operating frequency range \cite{Hul04}.  An Aharonov--Bohm ring uses magnetic flux to produce non-reciprocal transmission times in a simple  electron interferometer. In order to mimic that effect, we use a microwave gyrator \cite{Tell48,Hogan53,Viola14} to create a uni-directional $\pi$ phase shift in a microwave ring network.  As illustrated in Figs. \ref{gyrator} and \ref{phase_gyrator} this is achieved by means of two microwave circulators that are configured with short and open circuits on their third ports to create the requisite $\pi$ relative phase shift.  We also demonstrate in Appendix \ref{sec:ABMG} that this relative phase difference is approximately achieved over a broad frequency range (7-12.4 GHz) with virtually no difference in attenuation of the signals in the two directions. 

\begin{figure}[ht!]
    \centering
    \hspace*{-0.62cm}
    \includegraphics[width=0.46\textwidth]{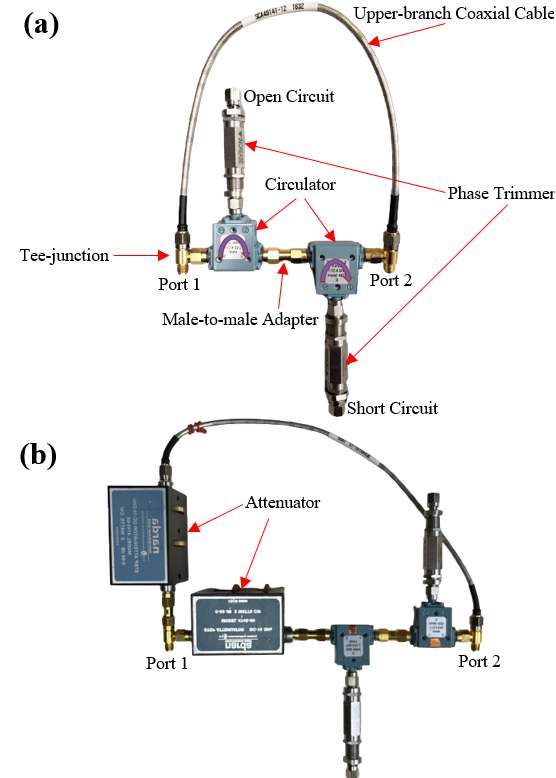}
    \caption{(a) Photograph of the Aharonov--Bohm microwave ring graph constructed with the gyrator from Fig. \ref{gyrator}. The upper branch is a single coaxial cable matching the electrical length of the lower branch. This is the experimental realization of the schematic shown in Fig. \ref{schematic_ABRing}(b), without the attenuators. (b) Microwave graph analogue Aharonov-Bohm ring for the two-attenuator case, with one attenuator of strength $\Gamma_A/2$ on each bond.}
    \label{Pics_ABRingGraph}
\end{figure}

Next, we construct the Aharonov--Bohm analogue ring microwave graph (see Fig. \ref{Pics_ABRingGraph}) using the gyrator design. The key is to add an upper branch to the ring that has the same electrical length as the gyrator, so that waves travelling from left to right in both branches would go through the same electrical length. Equating the electrical lengths of the branches also has the effect of eliminating the Feshbach modes, leaving only the shape resonances of the ring graph (see Refs. \cite{Chen2022} and \cite{Waltner2013}).  We measured the electrical length of the gyrator and a series of single coaxial cables, and selected a 12-inch-long coaxial cable as the upper branch. We then perform some fine-tuning on both phase trimmers to achieve an electrical length on the lower bond as close as possible to that of the 12-inch cable.
The 12-inch coaxial cable has an electrical length of $0.4386$ m, and we manage to adjust the electrical length of the gyrator to be $0.4385$ m while maintaining a near $\pi$ phase difference for left/right transmission. The resulting microwave analogue of the AB graph with $\pi$ flux-induced phase shift is shown in Fig. \ref{Pics_ABRingGraph}(a) (no lumped loss) and Fig. \ref{Pics_ABRingGraph}(b) (including lumped loss $\Gamma_A$).

The losses in the microwave graph arise from two sources.  There exists roughly uniform attenuation of the microwave signals associated with the insertion loss of the coaxial cables, circulators, phase trimmers, microwave tees, adapters, and terminations.  The attenuation $\eta$ of the coaxial cables used here, which dominates the uniform attenuation in the circuit, has been characterized in Appendix B of Ref. \cite{Chen2022}.  The second source of loss is associated with the variable lumped (localized absorbing channels) attenuators $\Gamma_A$, shown schematically in Fig. \ref{schematic_ABRing}(b).   
In the experiments both types of loss in the microwave graph will be considered.  Note
that in all comparisons of microwave data and mesoscopic theory we treat the independent variables frequency $f$ and energy $E$ as interchangeable.

\subsection{Frequency Domain Measurements}  \label{FDM} We measure the scattering ($S$) matrix of the 2-port microwave ring-graph as a function of frequency.  A Keysight N5242B network analyzer (PNA-X) is calibrated with a Keysight N4691D Electronic Calibration kit over the 7 to 12.4 GHz frequency range with a frequency step size of 168750 Hz.  An Agilent Technologies N5242A vector network analyzer (VNA) is also used over the same frequency range; it is calibrated with an Agilent N4691-60001 Electronic Calibration kit with the same frequency step size. The device under test (DUT) is attached at the calibration planes of the network analyzer and the $2\times 2$ $S$-matrix is measured as a function of frequency for various settings of the lumped attenuators in the microwave ring-graph.

To measure the response of the microwave ring-graph to incoherent input signals, a Microwave Semiconductor Corp (MSC) MC65242, 1.0 to 18 GHz noise source producing output excess noise ratio (ENR) of 31.5 dB, is used to create broadband noise in the microwave domain.  The noise signal is amplified by two Mini-Circuits amplifiers (ZX60-183A-S+) each with a bandwidth of  6 to 18 GHz.  The two amplifiers are separated by one Narda-MITEQ (94S46) attenuator with a bandwidth of 4 to 18 GHz. The amplified noise source is connected to one port of the microwave ring-graph while the other port of the graph is connected to a network analyzer (Keysight N5242B). The network analyzer is used in receiving mode, with a resolution bandwidth of 10 kHz and 200 point averaging, to measure the transmitted power in a frequency resolved manner.  The experiment is repeated with the ports on the microwave ring-graph reversed.  

\subsection{Time Domain Measurements}  The broadband nature of the 3:1 transmission time ratio of our AB-ring graph analogue allows us to utilize wide-bandwidth pulses to investigate the asymmetric transmission through the device. 
 We perform pulse propagation measurements through the microwave ring-graph in both directions, systematically varying the center frequency of the pulse, and measuring the time delays and amplitudes of the transmitted pulses.  A Tektronix model AWG70001B Arbitrary Waveform Generator (AWG) is attached to the input port of the DUT, and the other port is attached to a Keysight/Infiniium model UXR0104A real-time oscilloscope.  The input pulse created by the AWG is designed to be a 1-ns long Gaussian modulated pulse of center frequency $f_c$, with $f_c$ ranging in value from 7 to 12.4 GHz.  The pulse is measured by connecting the output of the AWG directly to the oscilloscope to establish the amplitude and timing of the incident pulse.  The measurement is then repeated with the microwave ring-graph present in the other orientation. 

We note that prior measurements of pulse transport in microwave graphs has focused on the delay distribution and identifying orbits due to short closed loops in the graph \cite{Sirko21}.

\subsection{Frequency-domain Experimental Results}

\begin{figure}[ht!]
    \centering
    \hspace*{-0.62cm}
    \includegraphics[width=0.58\textwidth]{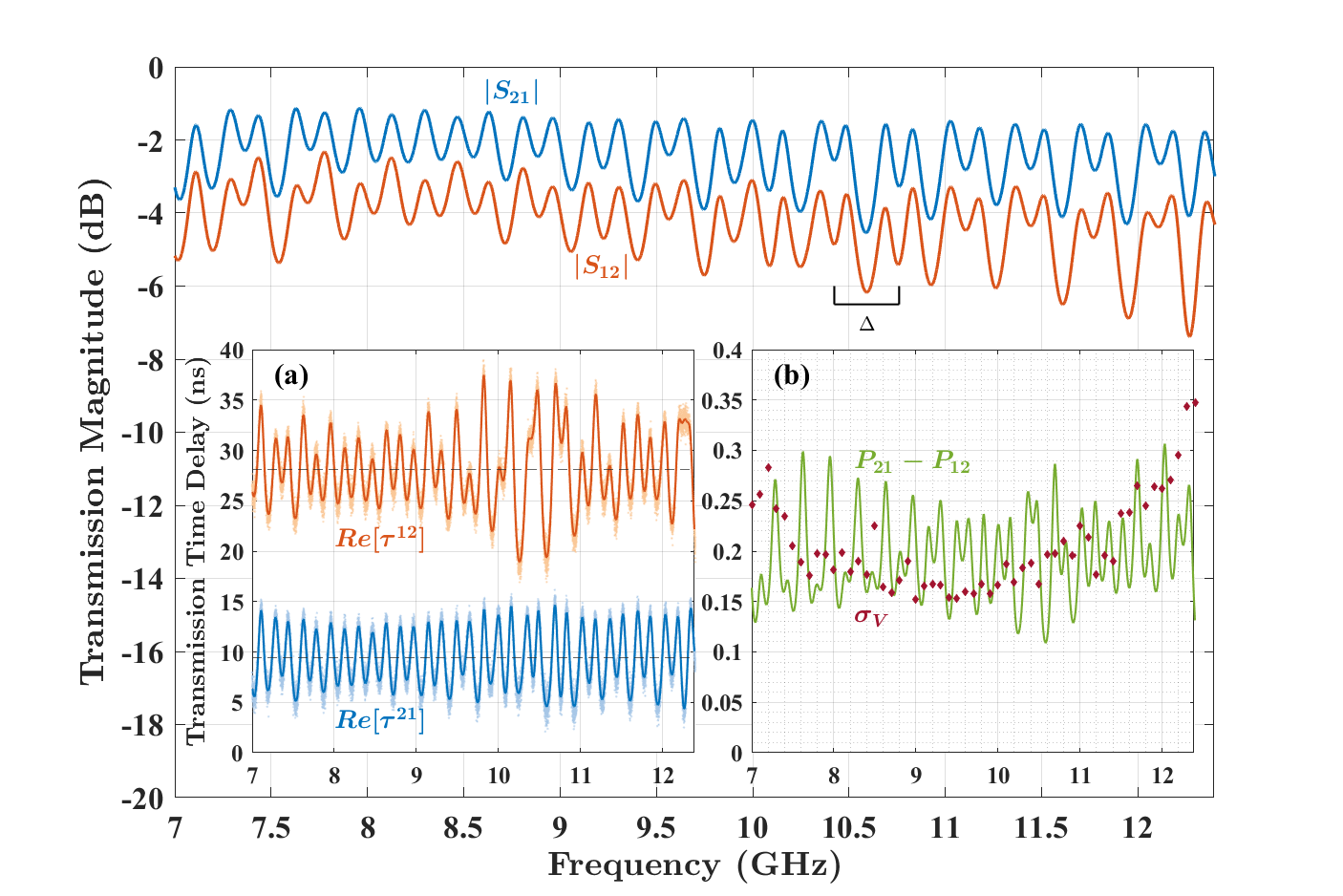}
    \caption{Measured $S$-matrix transmission data for the microwave ring-graph in Fig. \ref{Pics_ABRingGraph}(a) with approximately uniform attenuation, as a function of frequency. The scale bar gives the expected periodicity frequency scale ($\Delta = 0.342$ GHz) for the shape resonances of a simple ring graph. Note that $|S_{21}|$ is larger than the $|S_{12}|$ consistently across the full frequency range. Inset (a) shows the real part of the transmission time delays of the graph. We see that $\text{Re}[\tau_{T}^{12}] \approx 3\cdot \text{Re}[\tau_{T}^{21}]$ over the full frequency range. The dashed lines represent the average of the time delays. Inset (b) shows transmission probability asymmetry of the graph measured in both the frequency domain ($P_{21}-P_{12}$, green line) and time domain ($\sigma_V$, red diamonds) as a function of frequency.}
    \label{S_ABRingGraph}
\end{figure}

We measure the $2 \times 2$ scattering ($S$)-matrix of the Aharonov-Bohm ring microwave graph shown in Fig. \ref{Pics_ABRingGraph}(a) (with approximately uniform attenuation only) from 7 to 12.4 GHz, and plot the results in Fig. \ref{S_ABRingGraph}. The periodic wiggles in the plot come from the shape resonances of the graph, which have been thoroughly studied in Refs. \cite{Chen2022} and \cite{Waltner2013}, and discussed for the simulations. The periodicity of the shape resonances depends on the total circumferential length of the graph $\Sigma = 0.8771$ m, which corresponds to a repetition frequency of $\Delta = \frac{c}{\Sigma} = 0.342$ GHz. Note that waves propagating anti-clockwise around the microwave ring satisfy the ordinary shape resonance conditions, $f_n=\frac{c}{\Sigma} n$, with $n=1, 2, 3,...$.  However, waves travelling clockwise around the microwave ring suffer an additional $\pi$ phase shift upon passage through the gyrator, resulting in a different resonance condition, $f_m=\frac{c}{\Sigma} (m-\frac{1}{2})$, with $m=1, 2, 3,...$, creating a series of modes that begin at $\Delta/2 = 0.171$ GHz, and then alternate with the anti-clockwise modes as a function of frequency, consistent with the periodicity shown in Fig. \ref{S_ABRingGraph}.
This creates two distinct sets of S-matrix poles and zeros for the modes of the microwave ring-graph (see Supp. Mat. section ``Poles and Zeros of the S-Matrix of the Microwave Ring-Graph"). 

\subsection{Transmission Time Delays}

The measured results for the real part of transmission time delay in both directions, $\text{Re}[\tau_T^{21}]$, $\text{Re}[\tau_T^{12}]$, are shown in Fig. \ref{S_ABRingGraph}(a). The mean values are clearly different, while the regular variations of the transmission time delay plot come from the shape resonances of the ring graph, although here we are more interested in the ratio between the two transmission time delays. The mean values are $\braket{\text{Re}[\tau_T^{21}]} =$ 1.49 ns and $\braket{\text{Re}[\tau_T^{12}]} =$ 4.47 ns, with a ratio of 2.99.
This demonstrates the average 3:1 ratio for the two transmission time delays in the Aharonov--Bohm ring microwave graph, based on frequency-domain data.

\subsection{Asymmetric Transmission}

Figure \ref{S_ABRingGraph} shows a comparison between forward and reverse transmission amplitudes $|S_{21}|$ and $|S_{12}|$ in the microwave ring-graph, where it can be seen that $|S_{21}| > |S_{12}|$ for all frequencies in the bandwidth of the device. The asymmetric transmission probability $P_{21}-P_{12}$ is plotted as a function of frequency in Fig. \ref{S_ABRingGraph}(b) (green line), demonstrating the broadband nature of the effect.

\begin{figure}[ht!]
    \centering
    \hspace*{-0.6cm}
    \includegraphics[width=0.56\textwidth]{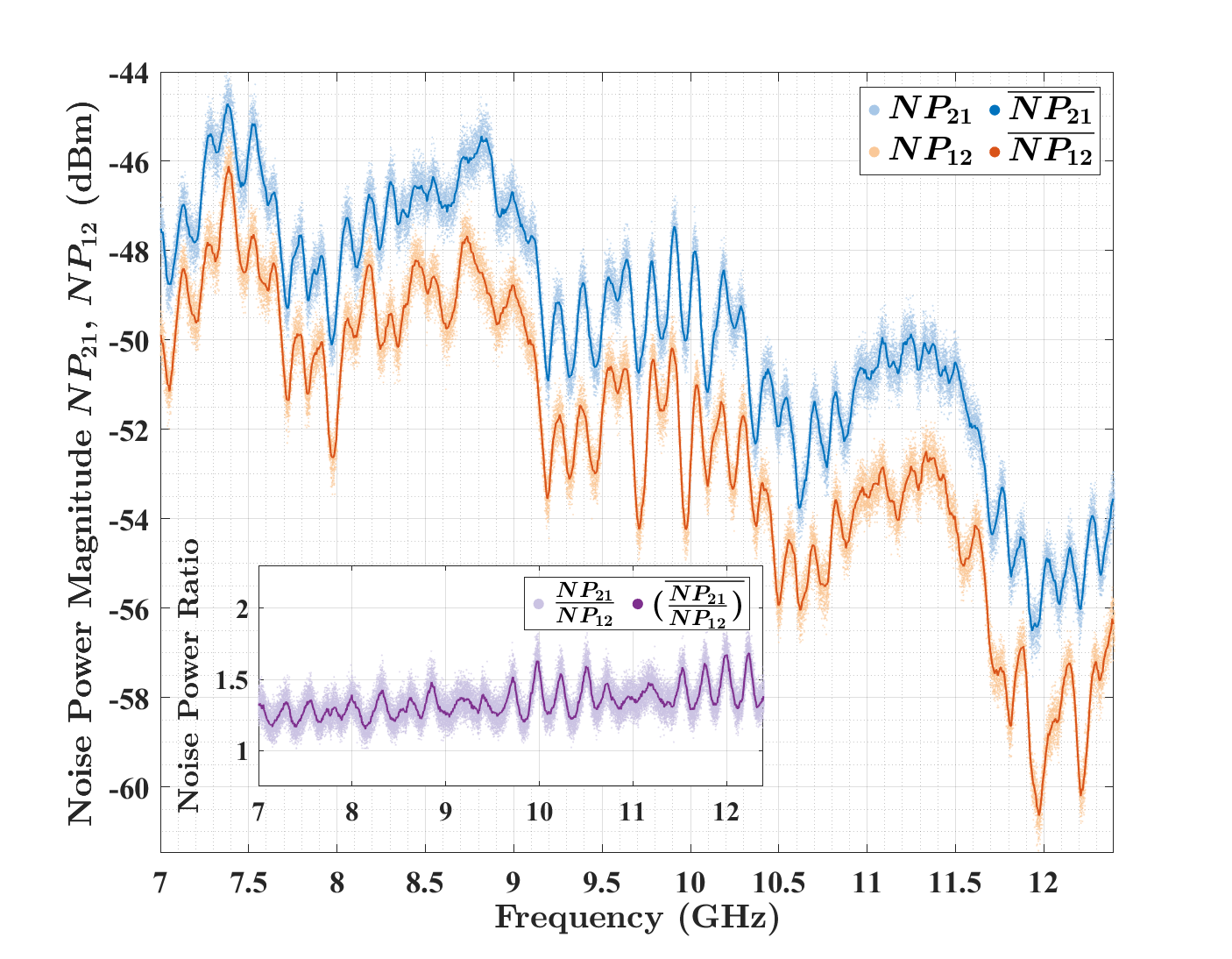}
    \caption{Measurements of asymmetric noise power transmission ($NP_{12}$ and $NP_{21}$ in units of $dBm$) as a function of frequency through the AB-microwave graph having only uniform attenuation, when subjected to a broadband noise source. Averaging ($\overline{NP}_{12}$, $\overline{NP}_{21}$) is performed by sectioning the data in groups of 50 frequency points and replacing each of these groups with their respective mean value. Inset shows the asymmetric transmission power ratio $\frac{NP_{21}}{NP_{12}}$ in linear scale over the bandwidth of the graph, showing a mean value of $1.35$. Averaging for $\overline{(\frac{NP_{21}}{NP_{12}})}$ is performed in the same way as described above for $\overline{NP}_{12}$ and $\overline{NP}_{21}$.}
    \label{fig:Noise}
\end{figure}

In addition to measuring the microwave ring-graph with coherent sources in the frequency domain and time domain, we are curious to see how it behaves when subjected to a broadband noise source.  It is interesting to see whether or not the asymmetric transmission properties are also exhibited when the AB-ring graph is excited by incoherent noise.  Figure \ref{fig:Noise} shows the results of such an experiment, in which a microwave noise source is used to illuminate one port of the graph.  In this case the transmitted signal is measured in a frequency-resolved manner by a network analyzer operating in receiver mode, as discussed in Section \ref{FDM}.
The experiment demonstrates broadband asymmetric transmission over the bandwidth of the device.  The bandwidth-averaged asymmetric transmission probability was measured as a function of total attenuation in the microwave ring-graph (Fig. \ref{Pics_ABRingGraph}(b)), and the results are shown in Fig. \ref{dephasing}(c).  This result shows that the non-monotonic dependence of transmission asymmetry exists even for a completely incoherent source of microwaves.

\subsection{\label{sec:TDE}Time-domain Experiments}
The results presented so far have been obtained from microwave scattering matrix measurements performed entirely in the frequency domain.  Electrons in mesoscopic systems are modeled as quantum wavepackets in the time domain.  Electron wave-packets do not have a single energy, as implied by our time-delay results obtain from S-matrix data, presented above. After demonstrating a 3:1 ratio between the two transmission time delays from the frequency-domain data, the question arises whether or not we can demonstrate the asymmetric delay directly in the time domain.  

\begin{figure}[ht!]
    \centering
    \hspace*{-0.62cm}
    \includegraphics[width=0.56\textwidth]{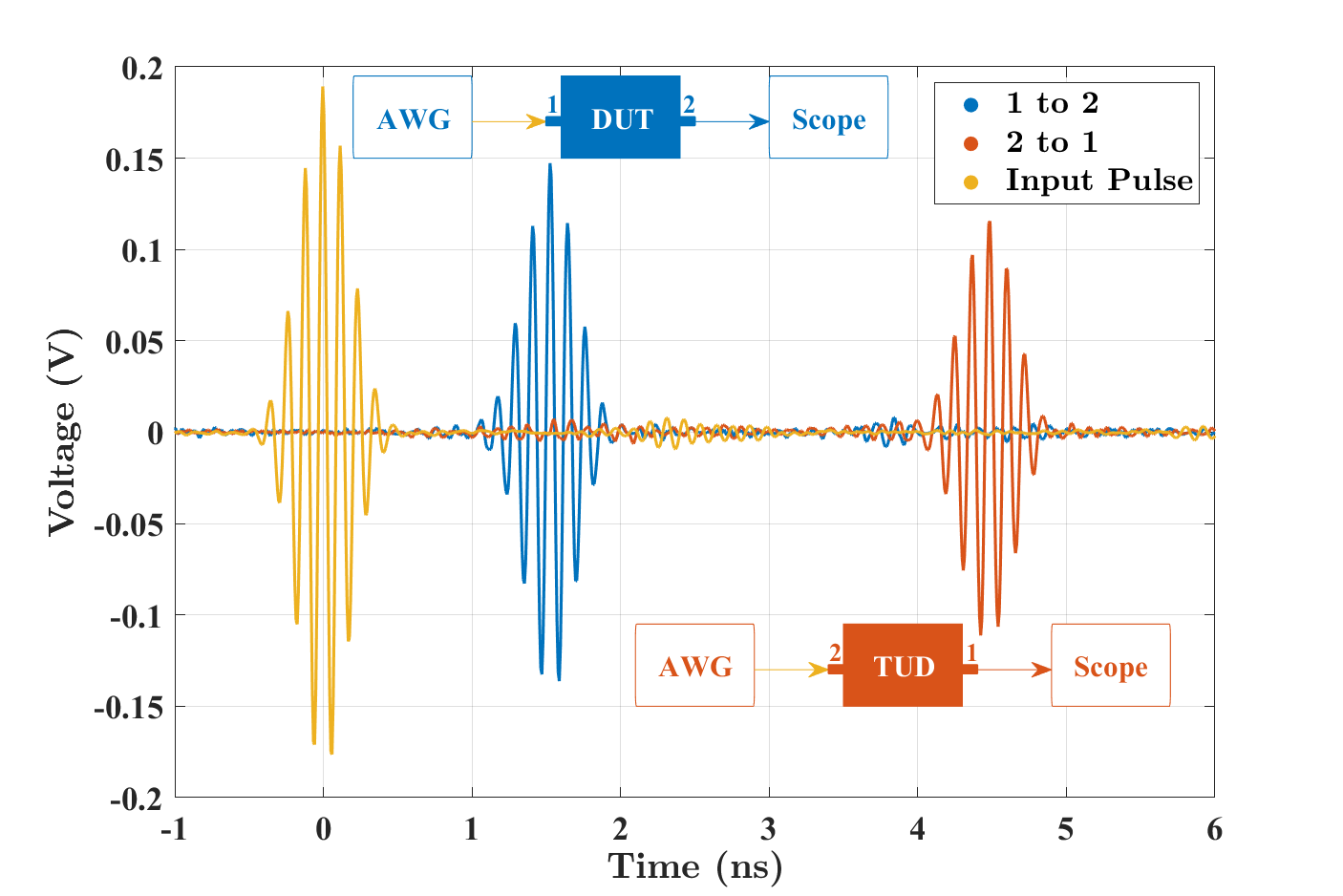}
    \caption{Time-domain measurements for the microwave ring-graph (the DUT here) shown in Fig. \ref{Pics_ABRingGraph}(a). Insets show the schematic of the time-domain setup for the $1\rightarrow 2$ and $2\rightarrow 1$ measurements. The yellow pulse is a measurement of the incident pulse (a 1-ns wide Gaussian amplitude modulation of a $f_c=8.5$-GHz carrier signal) from the arbitrary waveform generator (AWG) to the oscilloscope and establishes the baseline pulse amplitude. The blue pulse shows the transmitted pulse from port 1 to port 2, while the orange pulse shows the transmitted pulse from port 2 to port 1. Note that for the port 2 to port 1 measurement the DUT orientation is reversed.}
    \label{TimeDomain}
\end{figure}

Figure \ref{TimeDomain} shows representative time-domain results of transmitted pulses emerging at each port when sending in the pulse from the other port. Here, an arbitrary waveform generator (AWG) generates the gaussian-modulated pulse of fixed carrier frequency $f_c$ used in the time-domain measurements. The pulse is a 1-ns wide Gaussian amplitude modulation of an $f_c=8.5$ GHz carrier signal. Such a pulse includes approximately $1$ GHz bandwidth, which is well within the operating bandwidth of our gyrator-based microwave ring-graph. The pulse will have the effect of averaging the non-reciprocal properties of the graph over a finite bandwidth. Here DUT is the device under test (the microwave ring-graph), and the oscilloscope records the incident and transmitted signals. We measured the pulse transmission in both directions, and plot them on the same time axis. Shown for reference (yellow pulse in Figure \ref{TimeDomain}) is the incident pulse on the DUT, illustrating the relatively modest effects of pulse dispersion in the experiment. Figure \ref{TimeDomain} demonstrates directly a 3:1 ratio of the transmission time delays for pulses propagating through the AB graph in opposite directions. From the plot, the transmitted pulse from port 2 to port 1, $V_{12}$, has smaller amplitude compared to the transmitted pulse from port 1 to port 2, $V_{21}$. 

The red diamonds in Fig. \ref{S_ABRingGraph}(b) show the pulse transmission asymmetry parameter $\sigma_V = (|V_{21}|^2 -|V_{12}|^2)/|V_{in}|^2$ vs. pulse center frequency from the measured time-domain results in Fig. \ref{TD_FreqSweep}.  The results are in good agreement with the transmission probability asymmetry from frequency domain measurements of the same graph (green line).
Figure \ref{dephasing}(b) shows a plot of $\sigma_V$ at a pulse center frequency of 8.5 GHz vs.\ the lumped attenuation on each bond $\Gamma_A/2$, as they are varied together (red diamonds).  Adjusting both attenuators to the same attenuation setting preserves identical electrical path lengths on the two bonds of the microwave graph (Fig. \ref{Pics_ABRingGraph}(b)).  We see once again the non-monotonic dependence of transmission asymmetry on variable lumped attenuation, and good agreement between frequency domain $\braket{P_{21}-P_{12}}$ and time domain $\sigma_V$ measurements.







\section{Discussion}

\begin{figure}[ht!]
\includegraphics[width=0.49\textwidth]{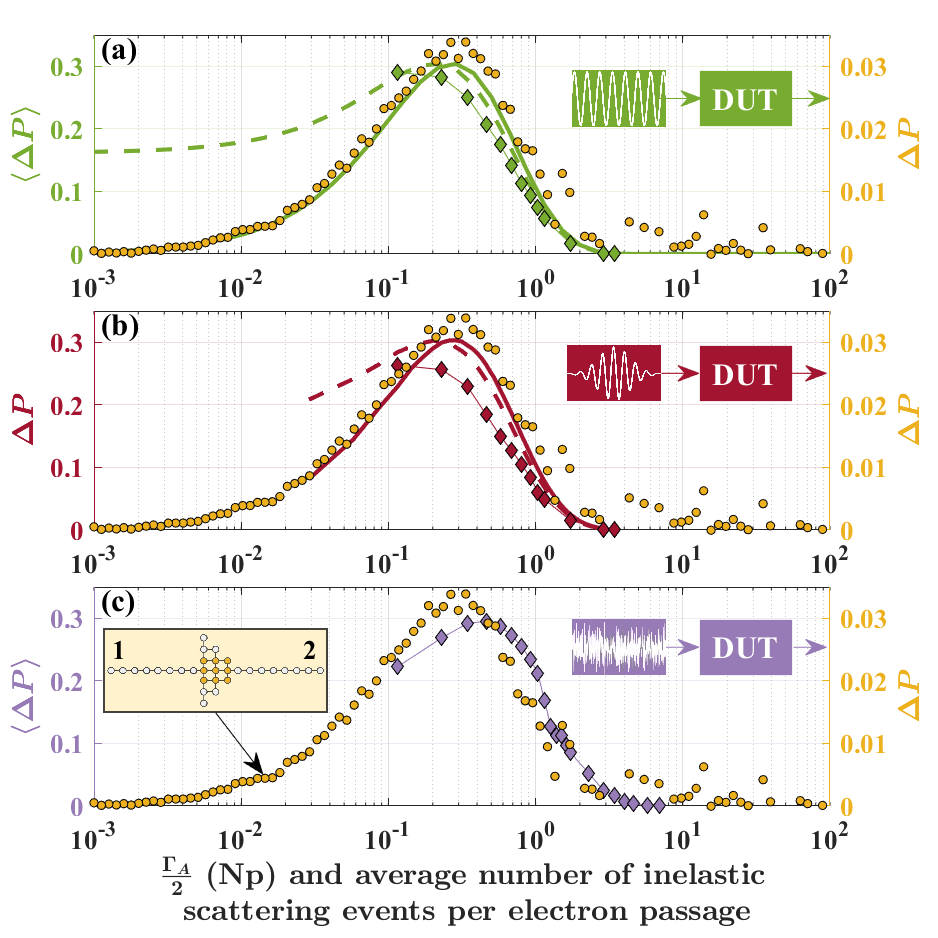}
\caption{Comparison of AB-graph microwave analogue asymmetric transmission ($\Delta P$ on left axis) and simulated mesoscopic device transmission probability asymmetry ($\Delta P$ on right axis), as a function of attenuation $\Gamma_A/2$ in Nepers, and average number of inelastic scattering events per electron passage, on a common log scale. Microwave measurements and simulation results are obtained in the (a) frequency-domain, (b) time-domain, and (c) frequency-resolved noise transmission.  Lines and symbols have the same meaning as those in the corresponding parts of Fig. \ref{dephasing}.  The mesoscopic simulation results $\Delta P$ (in yellow) are the same in all panels \cite{Bred21, Mann21}.}
\label{3-panel}
\end{figure}

We note some interesting parallels between the measurements on the classical microwave graph and earlier simulations of asymmetric transport in a model mesoscopic device with dephasing centers.  The mesoscopic calculation in Fig. \ref{dephasing}(d) shows a non-monotonic dependence of asymmetric left/right transmission probability $P_{21}-P_{12}$ as a function of the average number of inelastic scattering events per passage based on a phenomenological model of the quantum mesoscopic system, illustrated in the inset \cite{Bred21, Mann21}.  Plotting the lumped attenuation $\Gamma_A$ in units of Nepers (a $1/e$ decay of amplitude) is roughly analogous to the number of inelastic scattering events per passage of the wavepacket through the device.  A plot of asymmetric transmission vs. $\Gamma_A/2$ from microwave data has essentially the same dependence as simulated transmission probability asymmetry plotted vs. the average number of inelastic scattering events per passage, as shown in Fig. \ref{3-panel} (see also Fig. \ref{Sim_LogPlot_halfgamma} in Supp. Mat. section ``Detailed microwave AB-ring graph results").  The similarities are illustrated in Fig. \ref{3-panel} for (a) frequency-domain, (b) time-domain, and (c) broadband noise transmission.  In the unitary evolution (zero attenuation) case, there is symmetric transmission due to the purely coherent properties of the system.  At large de-coherence (attenuation) rates, the scattering events are so frequent that they destroy any bias for the electrons (microwaves) to follow a particular direction through the device, sometimes explained as a manifestation of the quantum Zeno effect \cite{Misra77}. Only in the intermediate de-coherence (attenuation) case does the combination of coherent transport, along with a finite degree of modeled de-coherence (attenuation) acting asymmetrically, result in a net transmission asymmetry through the device.  

From the perspective of the PCP, this unique state is not observed in either the purely coherent or strongly attenuated limits, but is a unique feature of wave systems augmented by a finite number of parasitic channels. 
Similar results on asymmetric transport of a Bose-Einstein condensate through an AB-ring analogue in cold atoms, as a function of loss rate (analogous the the attenuation used here), shows the same qualitative dependence as those in Figs. \ref{dephasing} and \ref{3-panel} \cite{ColdAtom20}.  It has also been proposed that transport of excitations across dissipative quantum networks can be enhanced by local dephasing noise, and the dependence of transport probability on dephasing rate is non-monotonic \cite{Plenio08,Reb09}.  Experiments on electronic excitation transport through a network of coupled trapped ions shows evidence of environment-assisted quantum transport (ENAQT) effects that are also a non-monotonic function of dephasing rate \cite{Maier19,Zerah20}. 
There are also proposals for dephasing assisted transport in coupled quantum dot systems \cite{Con14,Bett23}.

There have also been classical wave transport experiments that demonstrate similar behavior to that shown in Figs. \ref{dephasing} and \ref{3-panel}.  A classical optics experiment involving noisy resonators on a fiber network demonstrated a peak in the dependence of the transport efficiency as a function of the amount of `dephasing noise' in the network, realizing the phenomenon of `noise-assisted transport' \cite{Vici15}.  Another classical optics experiment utilizes variable input optical bandwidth to enhance transport \cite{Bigg16}.

It is remarkable that the measured microwave transmission asymmetry and the phenomenological mesoscopic dephasing treatment of asymmetric electron transport \cite{Mann21,Bred21} have the same dependence on the loss/dephasing parameter over such a wide dynamic range.  One possible reason for this is that the dephasing model effectively results in the same non-Hermitian treatment of the system Hamiltonian, and consequent modification of the scattering matrix, as observed for classical waves. This conclusion is consistent with all three observations discussed here, namely the experimental data on the classical wave analogue of the AB-graph, the simulated results for an asymmetric mesoscopic scatterer of Refs. \cite{Bred21, Mann21}, and the dissipative cold-atom experiment modeling an AB-graph in momentum space \cite{ColdAtom20}.  In fact, the non-monotonic dependence of transmission probability on `dephasing' or measurement rates seems to be a generic property of many treatments of quantum transport, such as environment-assisted quantum transport, \cite{Plenio08,Reb09,Maier19} and monitored quantum devices \cite{Annby22,ferreira2023exact}.  This leaves open the question of whether a more detailed or microscopic treatment of dephasing in quantum systems might arrive at different conclusions with regards to asymmetric transmission.

The microwave ring-graph elegantly captures the fundamental wave interference, non-reciprocal and dissipative properties that give rise to asymmetric transmission.  Of course, there are a number of limitations of the analogy between electromagnetic (EM) and mesoscopic quantum systems.  In the EM analogue there are no Fermi-Dirac statistics, a linear (as opposed to quadratic) dispersion relation, and no electron-electron interactions.  The physics of entanglement and superposition states are also absent in the classical case.  In the EM case the microwave photons are dissipated, whereas in the mesoscopic case the particle number is conserved.  Finally, the wave packet time evolution is different in detail, owing to the first-order nature of the time dependent Schrodinger equation, vs. the second order nature of the EM wave equation.  In addition the PCP approach to accounting for absorption in classical wave systems (or dephasing in quantum systems) is entirely phenomenological and generic in nature, detached from microscopic theory.  For example, it fails to capture multi-particle quantum interactions that lead to dephasing.

Next we comment on several properties of the microwave graph.
It is well established that non-reciprocal wave transport can be created by a simple microwave device known as an isolator.  One might ask: how does the microwave analogue of the AB-ring graph differ from an isolator?  An isolator can be fabricated from a three-terminal circulator, like those utilized in our design for the gyrator (see Fig. \ref{gyrator}).  The second port of the circulator is connected to a matched load that absorbs energy injected into port 1 and circulated to port 2.  Waves entering port 3 are directed to port 1, creating a two-port device that passes waves in one direction while absorbing those going in the opposite direction.  We note that photons absorbed and then re-emitted from the absorptive load will be forced to circulate to port 3, due to the design of the device.  In contrast the microwave AB-ring graph has a crucial difference.  Photons absorbed and re-emitted by the localized attenuators $\Gamma_A/2$ are able to exit the device through either port, with equal probability.  This feature is utilized in the phenomenological mesoscopic analogue model to create the asymmetric flow of electrons in that case \cite{Mann21,Bred21}.  

A second important feature of the microwave graph concerns its linearity.  It is well established that non-reciprocal transport can be achieved by exploiting nonlinearities.  A common example is the  diode, which has a nonlinear current-voltage characteristics at large voltage amplitudes, but a linear response at low amplitude \cite{BredSol21}.  In Fig. \ref{linearity} we demonstrate that the asymmetric transmission of the microwave AB-graph is present over a broad range of amplitude excitation.  It's ability to create asymmetric transmission is not based on nonlinear processes or features.

\section{Conclusions}
In this paper, we present a two-port microwave network structure that demonstrates a 3:1 ratio of the transmission time delays for waves travelling in opposite directions. The 3:1 ratio of the time delay has been illustrated in both frequency-domain and time-domain experiments. Using both simulations and experiments, we also demonstrate the asymmetric transmission through the microwave ring in both the frequency domain and time domain.  The degree of asymmetric transmission is shown to be a non-monotonic function of localized attenuation in the ring. The experimental results on asymmetric transmission exist for both coherent and incoherent signal sources.  We find that a particular phenomenological model of decoherence in mesoscopic quantum systems produces a very similar dependence of transmission asymmetry on dephasing rate.  



\bigskip
\textbf{Acknowledgements}  We acknowledge insightful discussions with Drs. Jochen Mannhart and Chris Jarzynski, and assistance from Jingnan Cai.  This work was supported by ONR under grant N000142312507, NSF/ECCS/RINGS under grant 2148318, DARPA/WARDEN under grant HR00112120021,  ONR/DURIP FY21 under grant N000142112924, and ONR/DURIP FY22 under grant N000142212263.


\bigskip


\appendix

\section{\label{sec:CST}CST Simulations}  Figure \ref{schematic_Att} shows the simulation schematic of the microwave ring-graph as a CST circuit model. Each port leads to a 3-way tee junction that acts as a beam-splitter and combiner for the incident waves from either direction, but having finite reflection for waves arriving on all three transmission lines.  The upper branch is a finite length of uniform coaxial cable transmission line, while the lower branch contains a model gyrator.  The upper and lower branches are chosen to have the same electrical length to eliminate Feshbach resonances of the ring \cite{Waltner2013,Chen2022} and both branches contain a lumped-loss variable attenuator.  The model closely approximates the experimental realization of the AB-analogue microwave graph.  The model allows introduction of lumped loss $\Gamma_A$ (variable attenuator) and uniform loss $\eta$ (in all the coaxial cables) to the graph. 

The shape modes of the ring graph involve standing wave patterns around the circumference of the ring and having a large overlap with propagating modes on the two leads \cite{Exner10,Waltner2013,Chen2022}.  The resulting poles and zeros of the scattering matrix are located relatively far from the real frequency axis \cite{Waltner2013,Chen2022}.  The Feshbach modes are an orthogonal set of modes that have minimal overlap with the extended modes on the leads, and create poles and zeros near the real frequency axis,\cite{Waltner2013,Chen2022} giving rise to narrow spectral features.

\section{\label{sec:ABMG}Aharonov-Bohm-Analogue Microwave Graph}  
\begin{figure}[ht]
\includegraphics[width=0.48\textwidth]{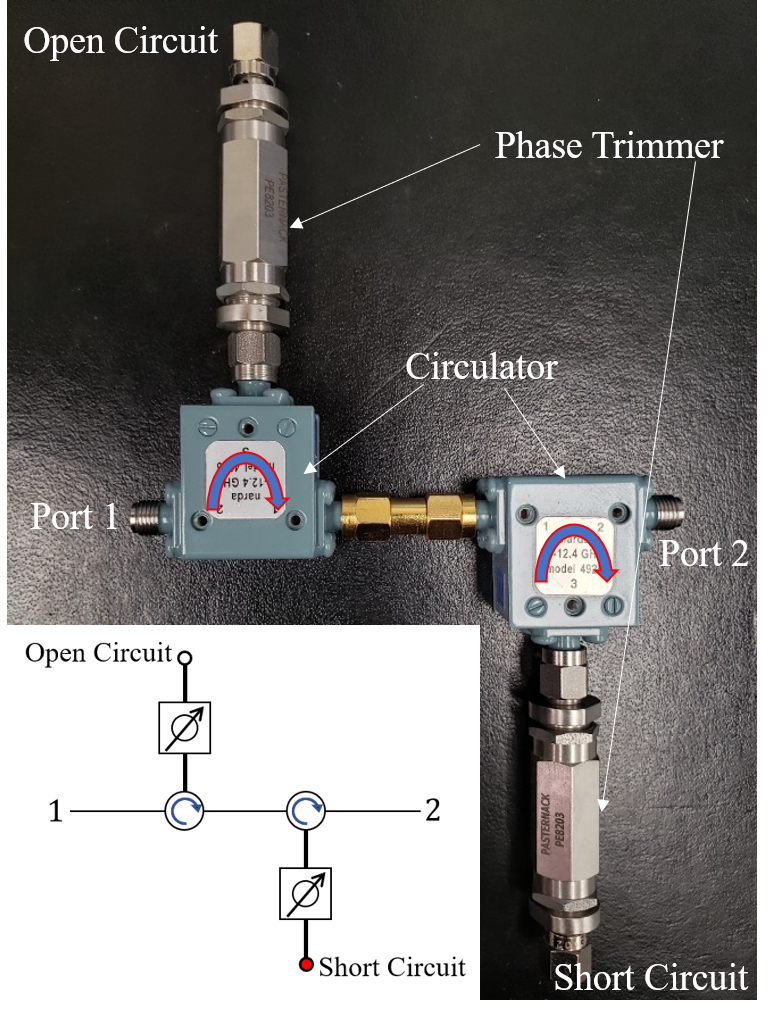}
\caption{The schematic design and photograph of a microwave gyrator offering non-reciprocal $\pi$ phase difference between the left-going waves and right-going waves traveling between ports 1 and 2. Two nominally identical circulators are used to guide the wave propagation directions, and a pair of open/short circuits are used to create the $\pi$ phase difference for waves travelling in different directions. Between the open/short circuits and the circulators, there are a pair of phase trimmers for calibrating the phase shift on the two vertical branches of the circuit.}
\label{gyrator}
\end{figure}

Figure \ref{gyrator} shows the schematic design and photograph of the microwave gyrator setup. A wave travelling from port 1 to port 2 will be directed upwards and reflect from the open circuit on the left circulator, while the wave travelling in the opposite direction from port 2 will be directed downwards and reflect from the short circuit on the right circulator.  The open/short circuits are designed to provide a $\pi$ phase difference for waves travelling in opposite directions, but due to the unequal finite electrical length of the components inside the circuits, the phase difference is not exactly $\pi$. Thus, we add a pair of carefully adjusted phase trimmers in the design to compensate for the difference in electrical lengths. (This basic idea of creating 0 and $\pi$ phase shifts for left and right-going waves was inspired by the graph presented in \cite{Rehemanjiang2016}) 

The measured phase difference between $S_{12}$ and $S_{21}$ between ports 1 and 2 in Fig. \ref{gyrator} as a function of frequency is shown in Fig. \ref{phase_gyrator}. The circulators used in the experiment have a working frequency range of $7-12.4$ GHz, and we measure the $S$-matrix of the gyrator in that frequency range. This broad frequency range enables the time-domain measurements with wave packets.  After doing some fine-tuning of the two phase trimmers, we are able to get the phase difference from the gyrator to be close to $\pi$ across the frequency band. We also verify that the insertion loss of the gyrator is symmetric (Fig. \ref{phase_gyrator}(b)).  There are some small wiggles in these plots, which are due to the imperfection of the circulators and resulting standing waves inside the circuit.  Note that the circulators contain microwave ferrites that are biased by a dc magnetic field in a fixed direction, dictating the circulation direction.  The operating bandwidth of the circulators is what limits the overall bandwidth of the entire microwave ring-graph.  Further characterization of the gyrator is given in the Supp. Mat. section ``Gyrator Properties."

Figure \ref{Pics_ABRingGraph}(b) shows a photograph of the assembled microwave AB-ring graph with two identical attenuators present.  Further characterization of this circuit is given in the Supp. Mat. section ``Detailed microwave ring-graph results."


\section{\label{sec:CTD}Complex Time Delay}  We have introduced a generalized version of time delay applicable to non-Hermitian (in this case sub-unitary) scattering systems, such as the ring graph with either uniform or lumped attenuation, or both.  The Wigner-Smith time delay for an $M$-port scattering system described by the $M\times M$ scattering matrix $S$ measured as a function of frequency $f$ is defined as \cite{Chen2021gen} $\tau_W \equiv \frac{-i}{M} \frac{d}{df}\log \det S(f)$.  This definition is a straightforward generalization of the Wigner-Smith time delay, previously considered only in the context of unitary quantum systems, to include variations in the magnitude of the scattering matrix with frequency.  The complex Wigner-Smith time delay as a function of frequency can be expressed in terms of the pole and zero locations of the scattering matrix $S$ \cite{Chen2021gen,Chen21} (see Supp. Mat. section ``Poles and Zeros of the S-Matrix of the Microwave Ring-Graph").  We also utilize the complex reflection and transmission time delays to identify the zeros of the reflection and transmission sub-matrices, respectively.  In this paper we evaluate the asymmetric complex transmission time delays of the $M=2$ port microwave ring-graph as \cite{Chen2022} $\tau_{T}^{12} \equiv -i \frac{d}{df} \log (S_{12}(f))$ and $\tau_{T}^{21} \equiv -i \frac{d}{df} \log (S_{21}(f))$.   We have previously utilized all of the complex time delays to characterize the shape and Feshbach modes of the ordinary (as opposed to AB-like) ring graph in terms of the zeros and poles of the scattering matrix and its transmission and reflection sub-matrices \cite{Chen2022}. 





\clearpage
\newpage


\pagebreak

\setcounter{figure}{0}
\setcounter{equation}{0}
\setcounter{section}{0}
\makeatletter
\renewcommand{\figurename}{Fig.}
\renewcommand{\thefigure}{S\arabic{figure}}   

\renewcommand{\tablename}{Table}
\renewcommand{\thetable}{S\Roman{table}}    

\renewcommand{\theequation}{S\arabic{equation}}    

\renewcommand{\thesection}{S\arabic{section}}
\makeatother


\maketitle
 
\date{\today}


\begin{center}

SUPPLEMENTARY MATERIAL 

\vspace{0.4cm}



Asymmetric Transmission Through a Classical Analogue of the Aharonov-Bohm Ring

\medskip
\vspace{0.4cm}
Lei Chen, Isabella L. Giovannelli, Nadav Shaibe, and Steven M. Anlage

\end{center}
\vspace{1cm}

Here we provide additional details for the experiments and simulations described in the text of the paper.  Also included are measurements of the poles and zeros of the scattering matrix of the AB-ring graph, and simulation and experimental results for an AB-ring graph with an attenuator on just one of the bonds.  Finally, we make some brief comments about dephasing models. \\

\begin{figure}[ht]
\includegraphics[width=0.48\textwidth]{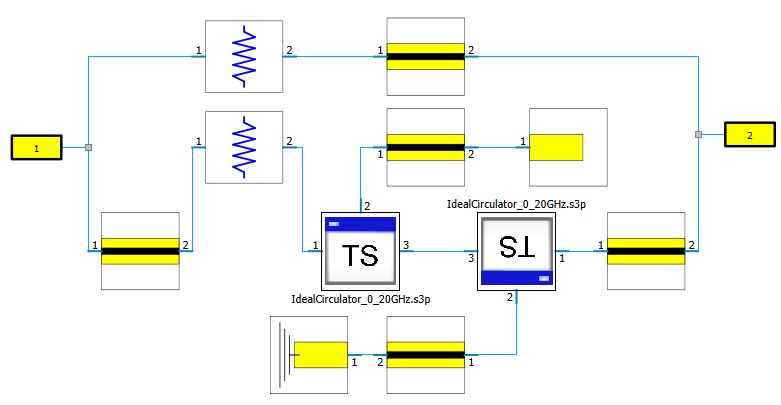}
\caption{Simulation schematic of the two-port microwave ring-graph in CST. A variable attenuator is included in both branches of the ring graph. The transmission line blocks represent the coaxial cables used in the experiment. The coaxial cables have two sources of uniform loss, i.e. the dielectric loss tangent of the medium $\tan\delta$, and the resistivity of the metals in the cable conductors, $\rho$. The open and short circuit blocks have a finite length of 0.013 cm with the same relative dielectric constant of 2.01. The two circulator blocks (denoted TS) are imported from TOUCHSTONE files as ideal circulator $S$-matrices that are constant over the range of $0-20$ GHz. The attenuators (boxes with resistor symbols) add a source of lumped loss.  The gyrator produces an asymmetric $\pi$ phase shift, giving the $\pi$ phase shift to the right-to-left travelling waves.  Note that the blue lines are zero-length connections between components while the grey squares are ideal 3-way tee junctions with finite reflection.  \label{schematic_Att}}

\end{figure}

\textbf{Simulations.}
Figure \ref{schematic_Att} shows the schematic of the CST model of the microwave AB-ring graph used for the simulations presented in the article. All the transmission line blocks (boxes with yellow and black stripes) represent homogeneous coaxial cables with inner diameter $a = 0.091$ cm, outer diameter $b =0.298$ cm, and relative dielectric constant $\epsilon_r= 2.01$. The coaxial cables have two sources of uniform loss, i.e. the dielectric loss tangent of the medium $\tan\delta = 0.00028$, and the resistivity of the metals in the cable $\rho = 4.4 \cross 10^{-3}$ $ \Omega$m. These values reflect the parameters of the physical coaxial cables used in the experiment. The frequency dependent uniform attenuation is given by (Appendix B of \cite{Chen2022}):
\begin{equation}
    \eta = \frac{1}{2}\left[2\pi f\tan\delta+\sqrt{\frac{2\pi f \rho}{2\mu_0}}\frac{1}{\sqrt{\epsilon_r}}\frac{1}{\ln{(b/a)}} \left(\frac{1}{a}+\frac{1}{b}\right)  \right]. \label{attneqn}
\end{equation}
The attenuator blocks (boxes with resistor symbol) act as lumped loss elements which have adjustable attenuation.

\begin{figure}[ht]
\includegraphics[width=0.48\textwidth]{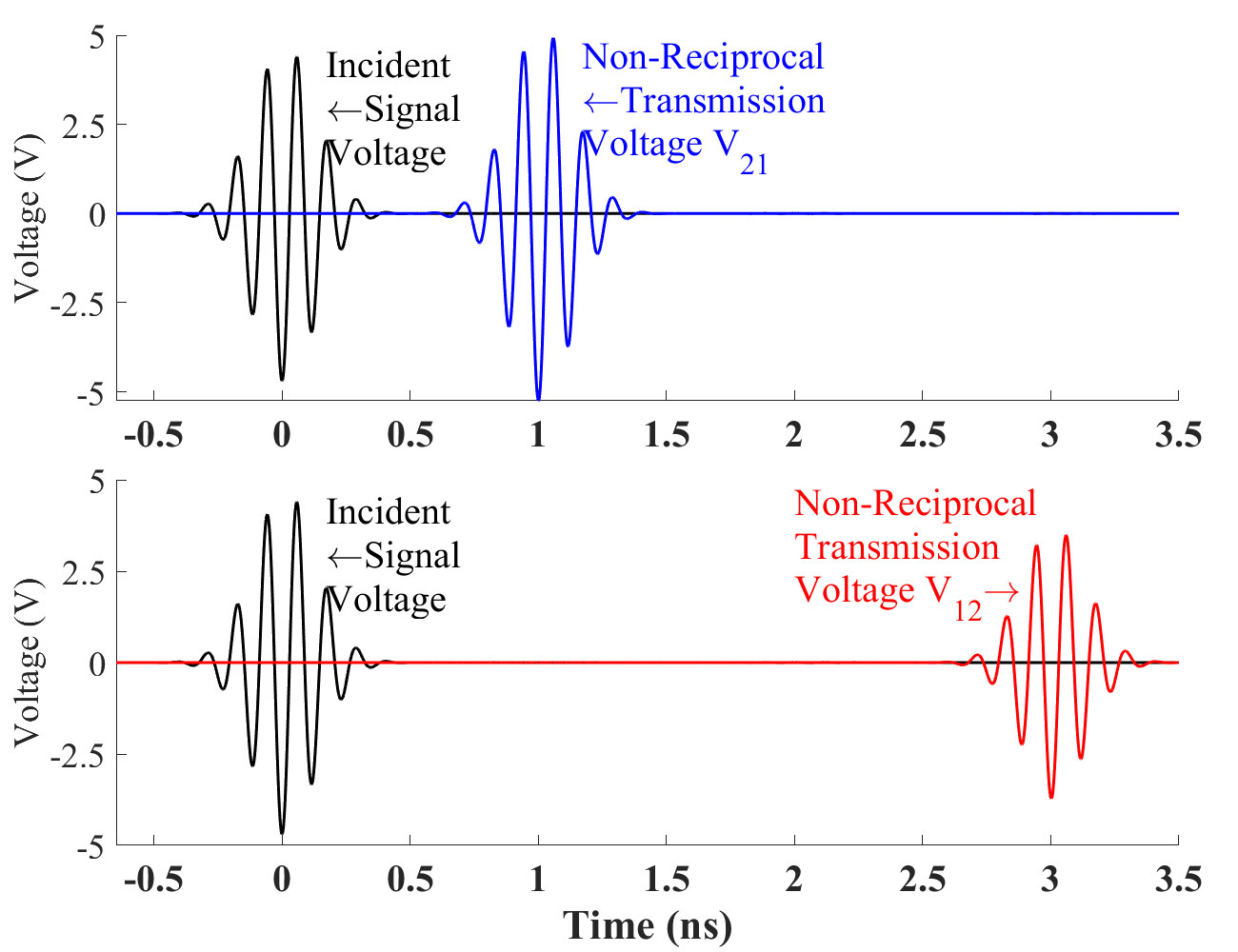}
\caption{Simulation of pulse propagation through the microwave ring-graph (Fig.\ \ref{schematic_Att}) in two different directions, conducted with both uniform attenuation and $0.35$ Np total lumped attenuation. Incident signal with center frequency $9.7$ GHz is displayed in black while the two transmitted pulses are displayed in blue for the signal traveling from port 1 to 2 (labeled $V_{21}$) and red for port 2 to 1 (labeled $V_{12}$). The 3:1 time-delay asymmetry is clearly evident, along with a significant difference in pulse amplitudes.}
\label{Sim_Pulse}
\end{figure}

For the simulation, we used a total graph length of $\Sigma = 0.6$ m, which corresponds to a repetition frequency of $\Delta = \frac{c}{\Sigma} = 0.5$ GHz. The reason for this choice is that each branch has a total length of $0.3$ m, so the transmission time through either branch is approximately $1$ ns. We can then most clearly display, as in Figure \ref{Sim_Pulse}, the 3:1 ratio of transmission times as in one direction it will be $1$ ns and in the other it will be $3$ ns due to the gyrator.

Both frequency domain and time domain simulations were done with this model. In the frequency domain, we investigated the range of $0-20$ GHz, but only show a small section of the full range in Fig. \ref{S_ABRingGraph_Att} 
since the behavior of $|S_{12}|$ and $|S_{21}|$ is largely repetitive in frequency. In the case of finite uniform attenuation $\eta$, both $|S_{12}|$ and $|S_{21}|$ decrease overall with increasing frequency as the total loss increases, and $|S_{12}|$ decreases faster as the waves travelling from port 2 to port 1 spend more time in the graph and therefore experience more loss.

For the time domain simulations, signals of varying duration and center frequency were put in the model as incident pulses. Figure \ref{Sim_Pulse} shows the propagation of a $1$ ns pulse with center frequency $9.7$ GHz through the microwave ring-graph in both directions. The incident pulse, portrayed in black, occurs at a normalized time of $0$ ns, and the transmission signal from port 1 to port 2 in blue occurs $1$ ns later, having the same temporal length and smaller amplitude. The transmission signal from port 2 to port 1, however, occurs at $3$ ns and has an even smaller amplitude. There is both uniform attenuation and a total of $0.35$ Np lumped attenuation in the graph in this case.

\begin{figure}[ht!]
\hspace*{-0.6cm}
\includegraphics[width=0.57\textwidth]{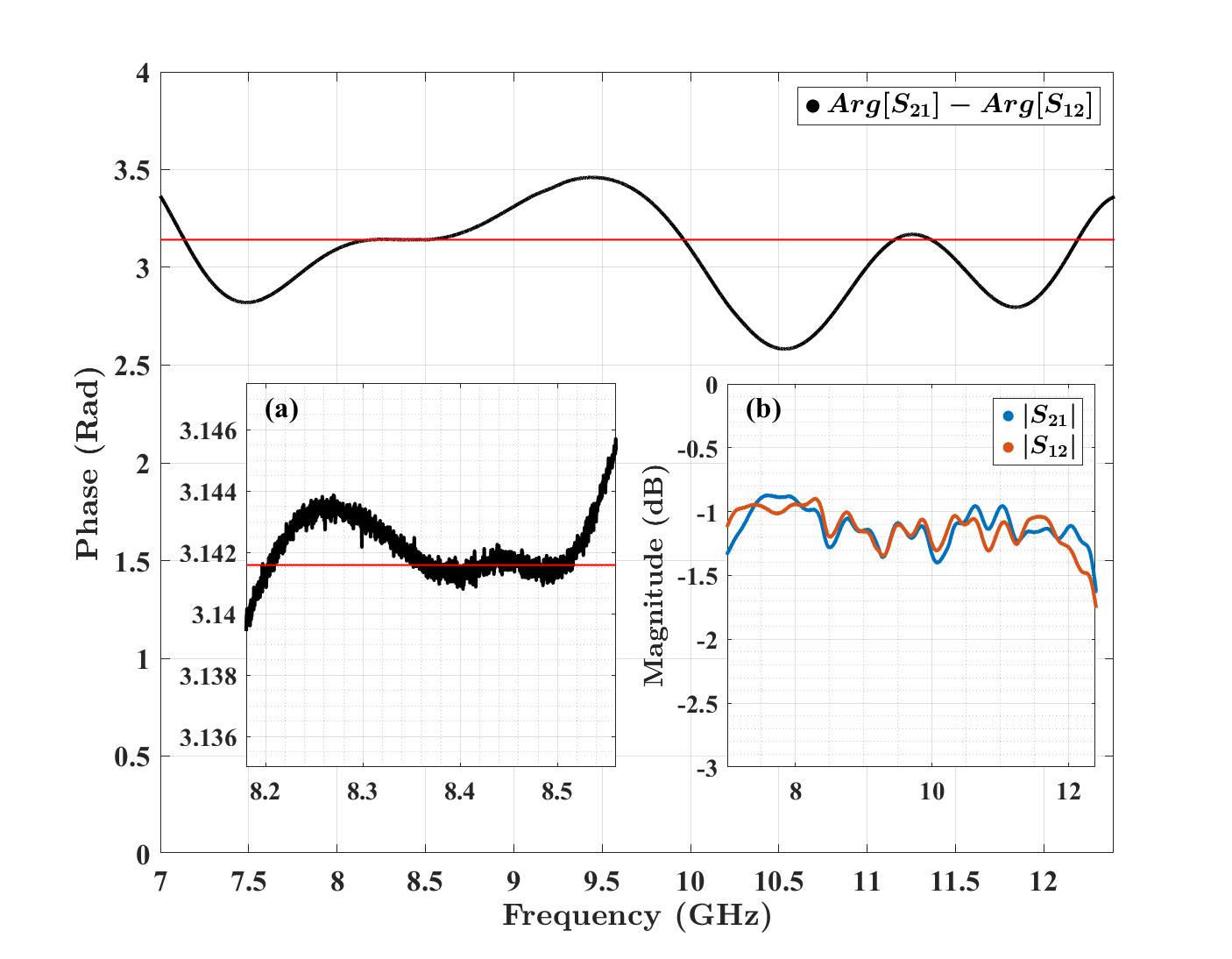}
\caption{Measured phase difference between $S_{12}$ and $S_{21}$ for the gyrator shown in Fig. \ref{gyrator}. The red line shows a reference phase difference of $\pi$. Inset (a) shows a zoom-in view of the plot. Inset (b) shows the comparison between $|S_{12}|$ and $|S_{21}|$ of the gyrator, demonstrating nearly equal insertion loss in both directions.}
\label{phase_gyrator}
\end{figure}

\textbf{Gyrator Properties.}
The gyrator schematic and laboratory realization are shown in Fig. \ref{gyrator}.  The measured transmission phase difference for right/left propagation through the experimentally realized gyrator is shown in Fig. \ref{phase_gyrator}.  Inset (a) of that figure shows a closeup of the phase difference between 8.2 and 8.5 GHz, where the device shows nearly ideal behavior.  Inset (b) shows the insertion loss for waves propagating in both directions, showing that they are nearly identical over the operating range of the gyrator.

Figure \ref{Pics_ABRingGraph}(a) shows a photograph of the assembled AB-ring microwave graph in the case where no lumped attenuators are present.  Figure \ref{Pics_ABRingGraph}(b) shows a photograph of the assembled microwave AB-ring graph with two identical attenuators present.

\textbf{Detailed microwave ring-graph results.}

\begin{figure}[ht!]
    \centering
    \hspace*{-0.6cm}
    \includegraphics[width=0.55\textwidth]{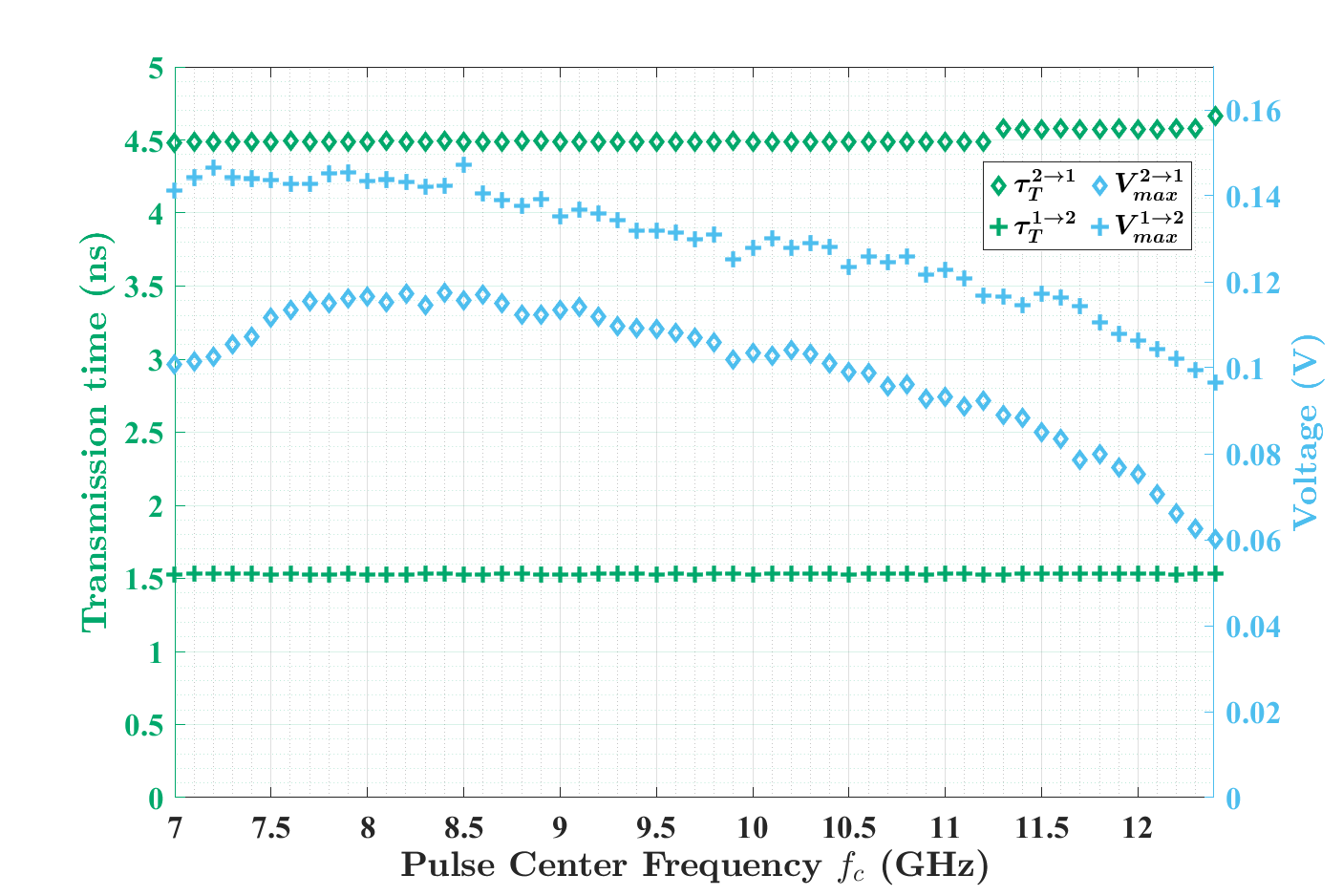}
    \caption{The two experimentally measured pulse transmission times and amplitudes as a function of center frequency $f_c$ of the 1-ns wide gaussian pulse, taken over the bandwidth of the microwave AB-graph device (7-12.4 GHz). The left axis (green) shows the two transmission time delays, while the right axis (blue) shows the two amplitudes of the transmitted pulses. The diamond points represent the measured data from port 1 to port 2, while the plus-sign points represents the measured data from port 2 to port 1.}
    \label{TD_FreqSweep}
\end{figure}

Figure \ref{TD_FreqSweep} shows a plot of the two measured transmission times and pulse amplitudes, as a function of center frequency of the 1-ns wide gaussian pulse, over the bandwidth of the AB-ring microwave device with uniform attenuation only (i.e. no lumped attenuation). These are the results discussed in section \ref{sec:TDE} and Fig. \ref{TimeDomain} of the main text.  The left axis (green) demonstrates a consistent 3:1 ratio of the transmission time delays for pulses with different center frequencies propagating through the microwave AB graph in opposite directions. The right axis (blue) shows that the transmitted pulse from port 2 to port 1 always has smaller amplitude compared to the transmitted pulse from port 1 to port 2. As the pulse center frequency increases, both amplitudes decrease because the uniform attenuation $\eta$ of the microwave graph increases at higher frequency \cite{Chen2022}, but the asymmetric amplitudes are maintained. (A similar dependence of transmission amplitude vs. frequency is seen in the noise power transmission data in Fig. \ref{fig:Noise}.)  The  resulting pulse transmission asymmetry $\sigma_V$ vs. center frequency for this time-domain data is shown in Fig. \ref{S_ABRingGraph}(b).

\begin{figure}[ht!]
\includegraphics[width=0.38\textwidth]{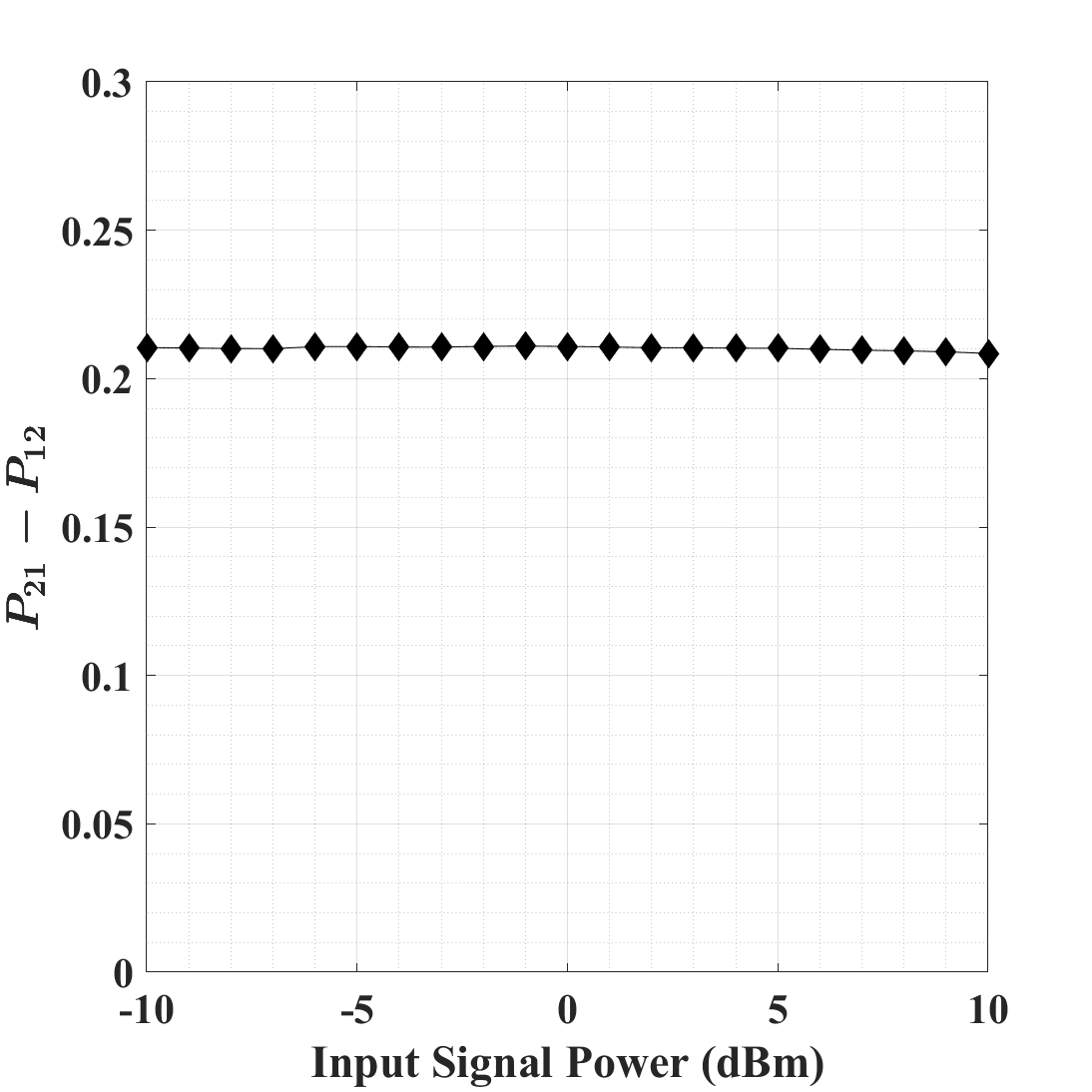}
\caption{Transmission probability difference $P_{21}-P_{12}$ through the microwave ring graph at 8 GHz versus input microwave signal power. This is a frequency domain measurement averaged over a 100 kHz bandwidth centered at 8 GHz.  The graph has no lumped attenuation. 
 The result demonstrates the linearity of the asymmetric transmission probability of the device.}
\label{linearity}
\end{figure}

Figure \ref{linearity} shows the measured asymmetric transmission of the microwave AB-ring with no lumped attenuation as a function of excitation power over a large dynamic range, from -10 dBm to +10 dBm.  The result illustrates that the microwave graph creates the asymmetric transmission in the linear-response regime.

\begin{figure}[ht!]
\includegraphics[width=0.48\textwidth]{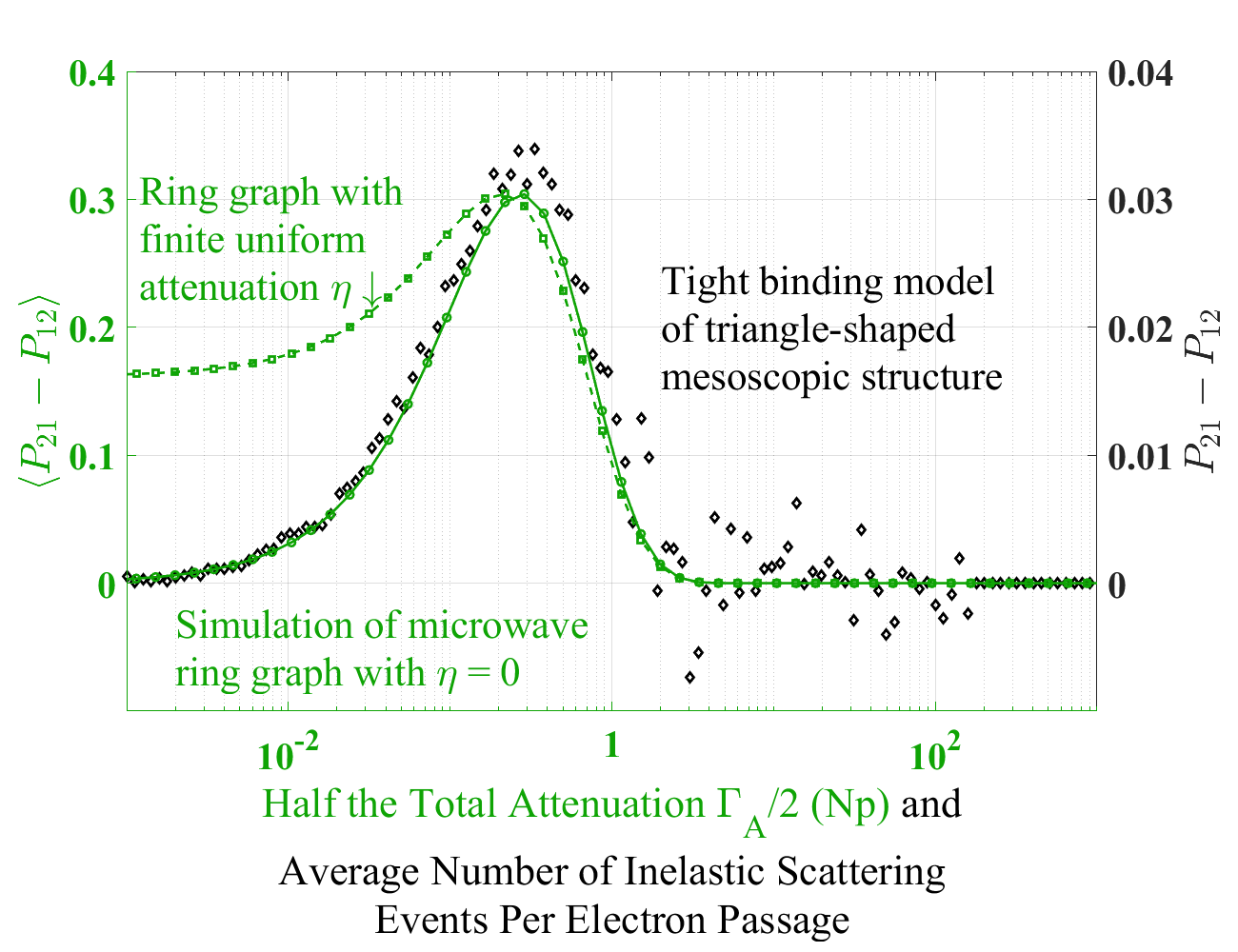}
\caption{Asymmetric transmission $P_{21}-P_{12}$ through the microwave AB-ring graph and a model mesoscopic device with similar asymmetric time delay. The green circles and squares (with y-axis on the left) correspond to asymmetric transmission $P_{21}-P_{12}$ of an Aharonov--Bohm microwave ring graph simulation with zero (green circles and solid line) and finite uniform attenuation $\eta$ (green squares and dashed line), respectively,  averaged over a period of the shape resonances, $8.25 - 8.75$ GHz, (denoted as $\braket{P_{21}-P_{12}}$) as a function of the lumped attenuation $\frac{\Gamma_A}{2}$ on each bond.  The black symbols (with y-axis on the right) correspond to the asymmetric transmission probability in a tight-binding model of a triangle-shaped mesoscopic device containing an extended model dephasing center, as a function of the average number of inelastic scattering events per electron passage through the device, from Refs. \cite{Mann21,Bred21}.  Data provided courtesy of Dr. Jochen Mannhart, Max Planck Institute for Solid State Research, Stuttgart, Germany.  Note the difference in y-axis scales. The x-axis is the same for both quantities.}
\label{Sim_LogPlot_halfgamma}
\end{figure}

Figure \ref{Sim_LogPlot_halfgamma} shows a plot of the simulated asymmetric transmission $P_{21}-P_{12}$ for both the simulation of the Aharonov--Bohm microwave ring graph and the tight-binding model of an asymmetric triangle-shaped mesoscopic electron device containing an extended dephasing center, from Refs. \cite{Mann21,Bred21}. Here the asymmetric transmission is presented as functions of total lumped attenuation in the microwave graph (in green) and average number of inelastic scattering events per electron passage through the modeled mesoscopic device (in black), respectively.  Note that both independent degrees of freedom are plotted on a logarithmic scale, as opposed to the linear scale used in Fig. \ref{dephasing}. Also note that both dimensionless rate quantities are plotted on the same scale.  The green circles and solid line are from a simulation with no uniform attenuation ($\eta=0$) while the squares and dashed line have a finite, frequency dependent uniform attenuation $\eta$.  Remarkably, the plot of the frequency averaged asymmetric transmission $\braket{ P_{21}-P_{12}}$ agrees in detail over many decades of attenuation/dephasing with the results of the mesoscopic, tight binding model calculations, although the vertical axes differ by a factor of 10.  The agreement for the degree of asymmetric transmission vs. attenuation/dephasing suggests a close connection between the two models for wave transport \cite{Sameer06}.

\textbf{Poles and Zeros of the $S$-Matrix of the Microwave Ring-Graph.}
The AB-ring graph resonance properties in the frequency domain can be understood from the locations of the poles and zeros of the scattering matrix in the complex frequency plane. It is known that uniform attenuation preserves the complex conjugate relationship of the $S$-matrix zeros and poles for each mode, but adds an overall offset $f \rightarrow f + i\eta$ in the definition of the $S$-matrix \cite{Chen2021gen, Chen2022}.  The lumped loss $\Gamma_A$ will act on the poles and zeros differently, allowing the real and imaginary parts of the poles and zeros to vary independently \cite{Chen2021gen}.  In particular, the zeros of the $S$-matrix can move across the real frequency axis and go into the lower-half of the complex frequency plane \cite{Erb23}.

\begin{figure}[ht!]
\includegraphics[width=0.48\textwidth]{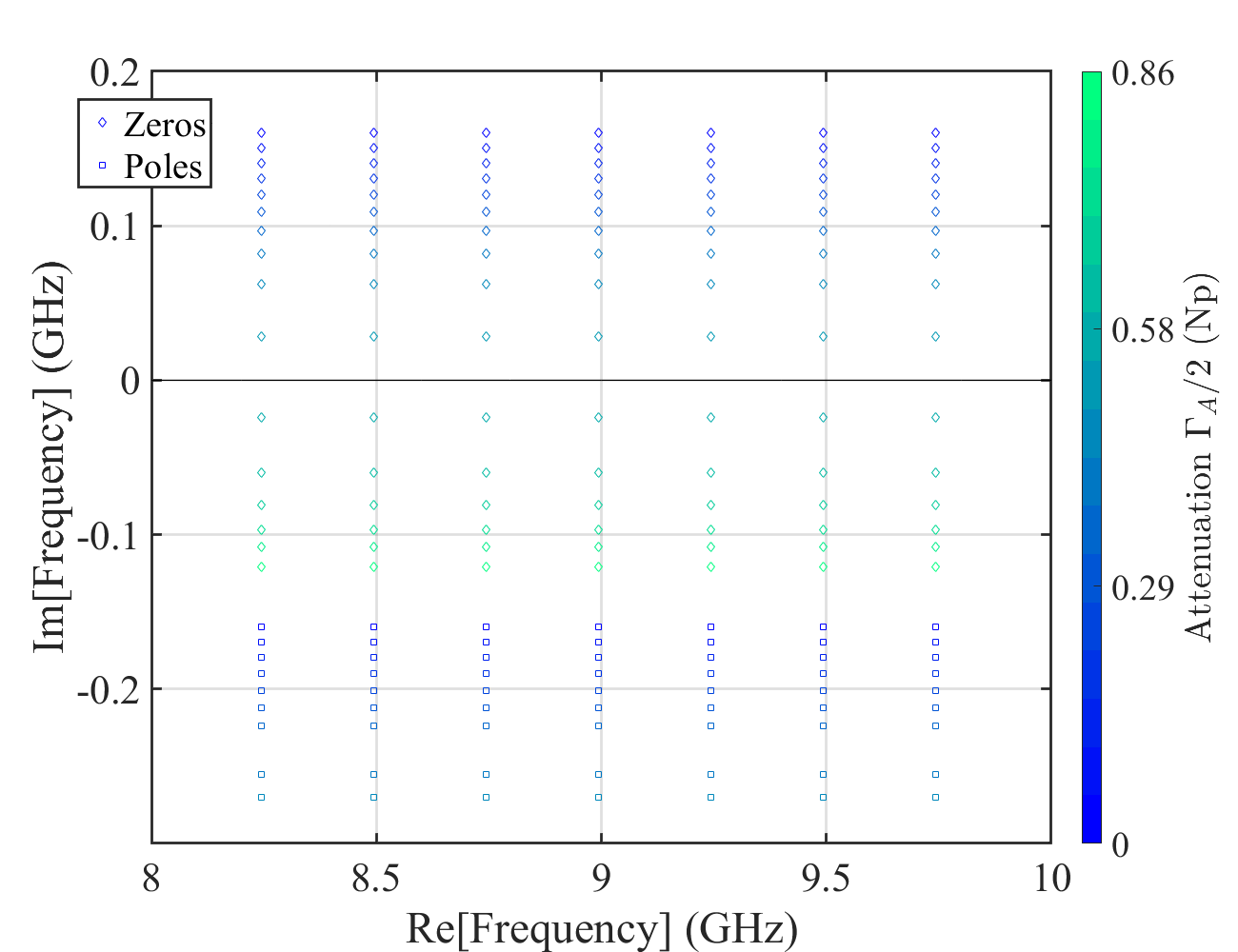}
\caption{Locations of the zeros (diamonds) and poles (squares) of the $S$-matrix in the complex frequency plane from the AB-ring graph simulation in the case of no uniform attenuation ($\eta=0$) and variable lumped attenuation $\Gamma_A/2$.  The zeros and poles are complex conjugates of each other when the lumped loss $\Gamma_A=0$.  As the total attenuation on the two bonds increases, represented by the color shift from blue to green (color bar on the right), the imaginary components of both the zeros and poles decrease while the real components remain unchanged.  The zeros cross the real frequency axis at $\Gamma_A/2 \approx 0.52$ Np.  A representative frequency range including 7 modes is shown.}
\label{Sim_PnZ}
\end{figure}

Figure \ref{Sim_PnZ} shows the motion of the zeros and poles (represented by diamonds and squares, respectively) as the lumped loss $\Gamma_A/2$ is increased from $0$ to $0.86$ Nepers in the simulation of the AB-like microwave ring graph with $\eta=0$.  The zero and pole locations are obtained by fitting the Wigner-Smith time delay $\tau_W$ as a function of frequency.  The complex Wigner-Smith time delay is defined as  $\tau_W \equiv \frac{-i}{M} \frac{d}{df}\log \det S(f)$, where $M$ is the number of ports, and is calculated from the measured (or simulated) scattering matrix in the frequency domain.  These frequency dependent functions are simultaneously fit to these expressions \cite{Chen2021gen,Chen2022}:

\begin{multline}
\text{Re}\: \tau_{W} = \frac{1}{M} \sum\limits_{n=1}^{N} [ \frac{\text{Im} [z_n] - \eta}{(f-\text{Re} [z_n])^2 + (\text{Im} [z_n] - \eta)^2} + \\
\frac{\Gamma_n + \eta}{(f-f_n)^2 + (\Gamma_n + \eta)^2}]
\label{Retau}
\end{multline}
\begin{multline}
\text{Im}\: \tau_{W} = -\frac{1}{M} \sum\limits_{n=1}^{N} [ \frac{f-\text{Re} [z_n]}{(f-\text{Re}[z_n])^2 + (\text{Im} [z_n] - \eta)^2} - \\
\frac{f-f_n}{(f-f_n)^2 + (\Gamma_n + \eta)^2}]
\label{Imtau}
\end{multline}
where the complex frequency locations of the poles ($\varepsilon_n =  f_n -  i \Gamma_n$, with the convention that $\Gamma_n>0$ in passive lossy systems) and zeros ($z_n =  \text{R}e[z_n] + i \text{Im}[z_n]$) of each mode $n$ are used as fitting parameters.  As before, $\eta$ is the uniform attenuation. 
 For the simulations used to generate $S$-matrix data that is analyzed in Fig. \ref{Sim_PnZ}, the increments in lumped attenuation are uniform, but the changes in the imaginary components of the zeros and poles are not, with the changes becoming more pronounced as the zeros approach the real axis.  Interestingly, the zeros cross the real axis at around $\Gamma_A/2 = 0.52$ Np, while the peak in asymmetric transmission $\braket{P_{21}-P_{12}}$ occurs at a little under $\Gamma_A/2 = 0.3$ Np, suggesting the two events may be related. Note that the fits to complex $\tau_W$ are dominated by the zero locations when they are in the vicinity of the real frequency axis, hence the pole locations are not determined for $\Gamma_A/2 > 0.45$ Np.

\begin{figure}[ht]
\includegraphics[width=0.48\textwidth]{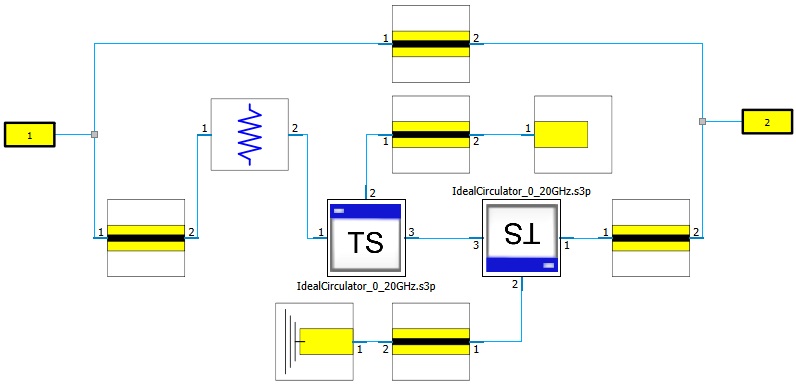}
\caption{Simulation schematic of the microwave ring-graph in CST. This differs from Fig. \ref{schematic_Att} as a variable attenuator $\Gamma_A$ is added to only the branch with the gyrator. That is the only change between the two schematics.}
\label{schematic_1Att}
\end{figure}

\begin{figure}[ht!]
\hspace*{-0.8cm}
\includegraphics[width=0.55\textwidth]{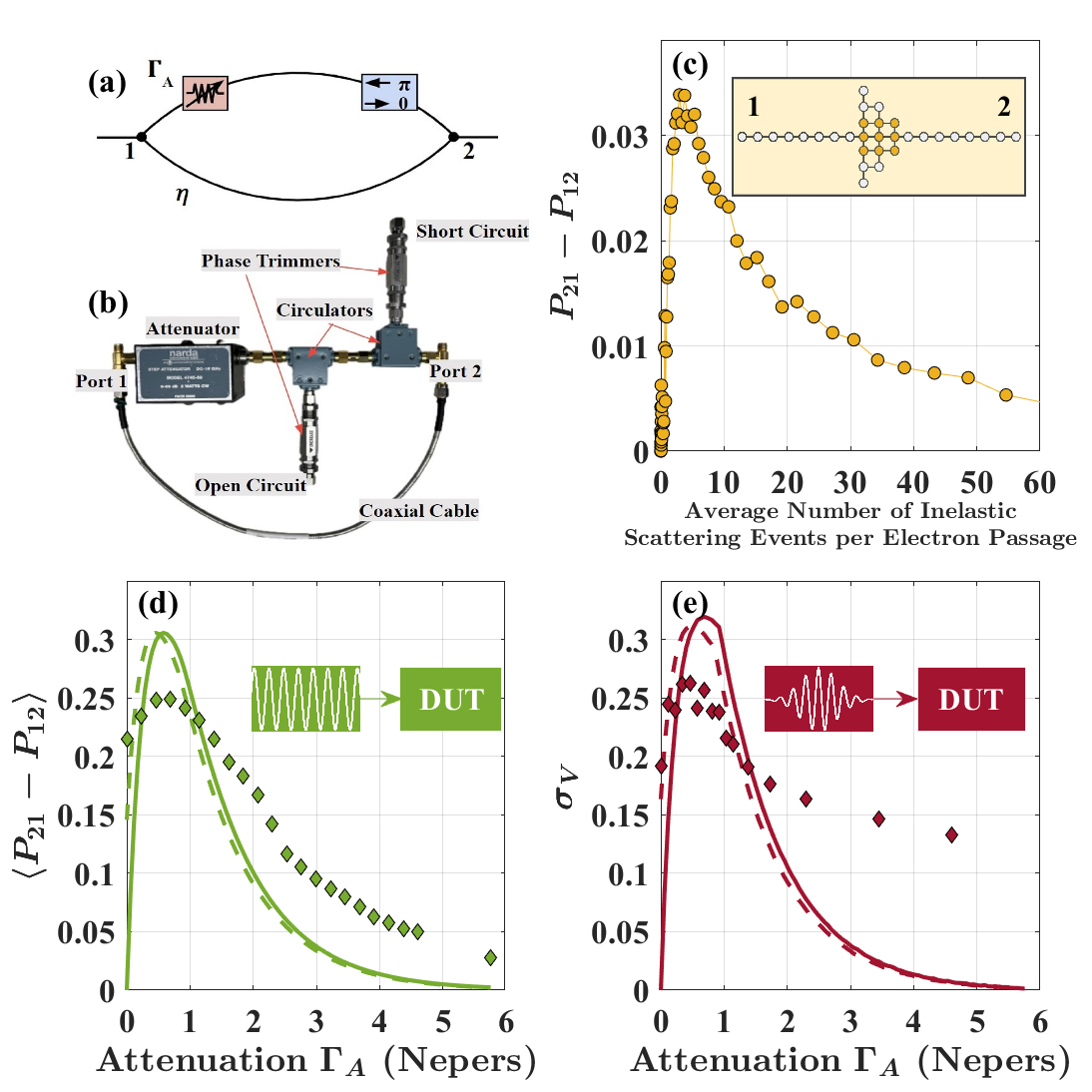}
\caption{Asymmetric transmission through an AB-analogue microwave ring graph in the single attenuator case, shown schematically in Fig. \ref{schematic_1Att}.  (a) Shows a schematic of the AB-ring graph with only one attenuator on one bond.  (b) Shows the experimental realization of the schematic in (a).  (c) Asymmetric transmission probability in a tight-binding model of triangle-shaped mesoscopic device with a model dephasing center, as a function of the average number of inelastic scattering events per electron passage through the structure.  From Refs. \cite{Mann21,Bred21}.  Data provided courtesy of Dr. Jochen Mannhart, Max Planck Institute for Solid State Research, Stuttgart, Germany. (d) Summary of the frequency domain asymmetric transmission results. For all curves, the quantity $P_{21}-P_{12}$ is averaged over the frequency range $8.25 - 8.75$ GHz, which is denoted as $\braket{P_{21}-P_{12}}$. The solid line is from simulation where only the localized attenuation from the single attenuator is considered. The dashed line is also from simulation but for the more realistic case where finite uniform attenuation throughout the device is taken into account. The discrete diamond points are from experimental data. (e) Summary of time domain results. For these time domain measurements, a 1 ns wide Gaussian pulse with a center frequency of 8.5 GHz is sent into the device. The solid line is from simulation where only the localized attenuation from the single attenuator is considered. The dashed line is also from simulation but for the more realistic case where finite uniform attenuation throughout the device is taken into account. The discrete diamond points are from experimental data.}
\label{1Atten}
\end{figure}

\textbf{Single-attenuator microwave AB-ring graph.}
A version of the AB-ring graph was also constructed with a single lumped attenuator on the lower bond, the schematic of which is shown in Figure \ref{schematic_1Att}.  This version of the graph is meant to be a closer analogue of the schematic form shown in Fig. \ref{schematic_ABRing}(a).  A photograph of the experimentally realized graph is shown in Fig. \ref{1Atten}(b).  The un-balanced nature of the attenuation in the two bonds leads to more complicated behavior of the scattering properties as the attenuation value is increased.

\begin{figure}[ht]
\includegraphics[width=0.48\textwidth]{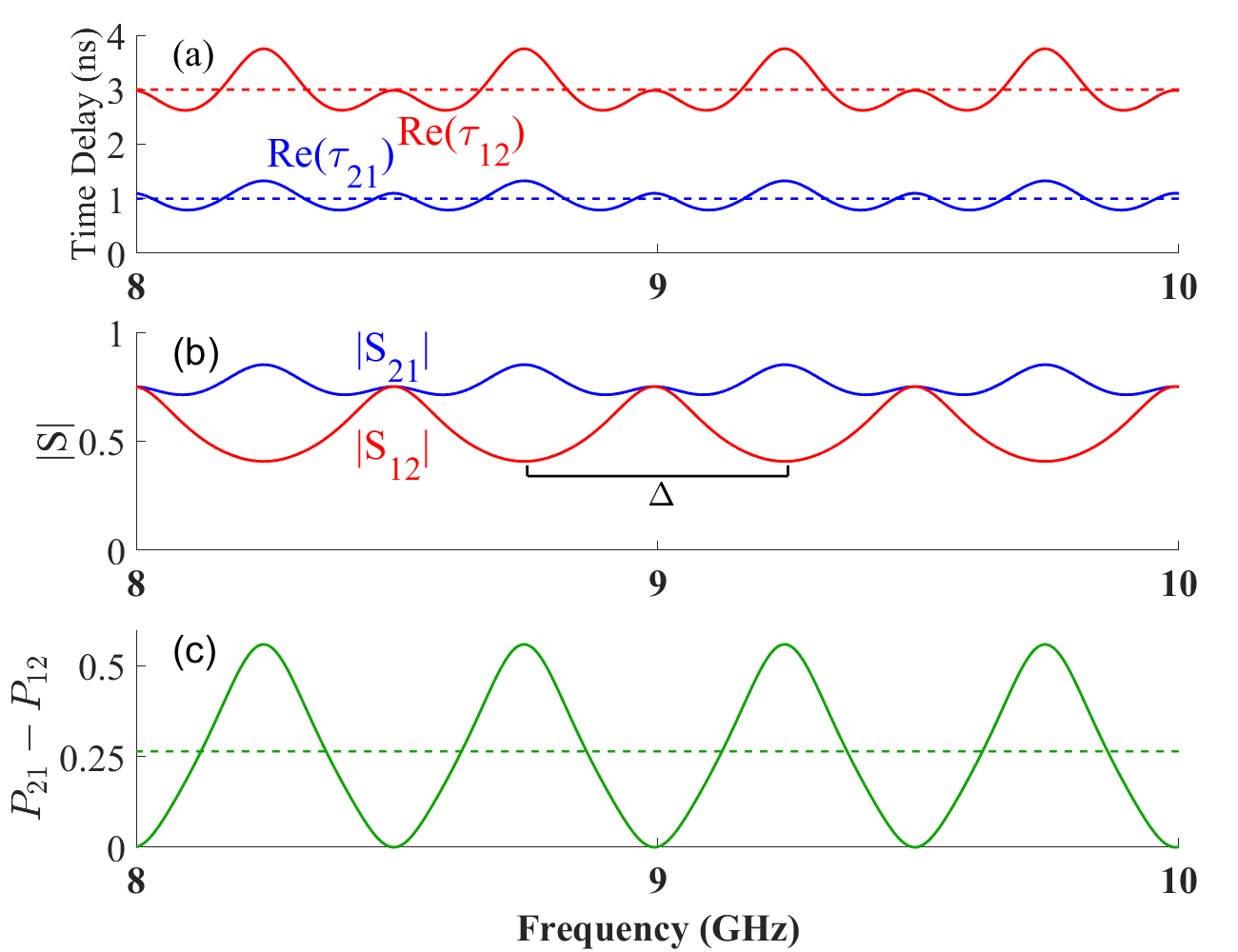}
\caption{(a) Real part of transmission time delays of the Aharonov-Bohm microwave ring graph simulation for the case of $0.35$ Np attenuation on only the bottom branch and uniform attenuation throughout (see the schematic in Fig. \ref{schematic_1Att}). Blue curve shows $\text{Re}[\tau_{21}]$ and red curve shows $\text{Re}[\tau_{12}]$, and the dashed lines show the mean values, demonstrating a 3:1 ratio. (b) Linear transmission magnitudes in both directions for the case of $0.35$ Np unbalanced attenuation. Blue corresponds to $|S_{21}|$ while red corresponds to $|S_{12}|$. With no attenuation the transmission magnitudes are identical. The scale bar gives the expected periodicity frequency scale ($\Delta = 0.5$ GHz) for the shape resonances of a ring graph. (c) The asymmetric transmission probability $P_{21}-P_{12}$ for the case of $0.35$ Np unbalanced attenuation. The $0$ Np case was not displayed in panels (b) and (c) as it is the same as that shown in Figure \ref{S_ABRingGraph_Att}.}
\label{S_ABRingGraph_BottomAttn}
\end{figure}

Figure \ref{S_ABRingGraph_BottomAttn}(a) shows the simulated real part of the complex transmission time delay through the un-balanced version of the AB-ring graph as a function of frequency.  The 3:1 asymmetry in transmission time delays is clearly seen, along with variations associated with the strong shape resonances.  Figure \ref{S_ABRingGraph_BottomAttn}(b) shows the simulated forward and reverse transmission amplitudes as a function of frequency for the case of lumped attenuation $\Gamma_A = 0.35$ Np, with zero uniform attenuation.  The case of no loss is not included here as it is identical to having two attenuators set to $0$ Np, shown in Figure \ref{S_ABRingGraph_Att}, resulting in the graph having a unitary scattering matrix and symmetric transmission.  In the $0.35$ Np lumped attenuation case the transmission is asymmetric for all frequencies except discrete values that repeat at intervals of the periodicity $\Delta = c/\Sigma = 0.5$ GHz.  Figure \ref{S_ABRingGraph_BottomAttn}(c) shows the difference in simulated transmission probabilities through the graph as a function of frequency in the unbalanced $0.35$ Np case, while the $0$ Np case has $\langle P_{21}-P_{12} \rangle=0$ at all frequencies.

\begin{figure}[ht]
\includegraphics[width=0.48\textwidth]{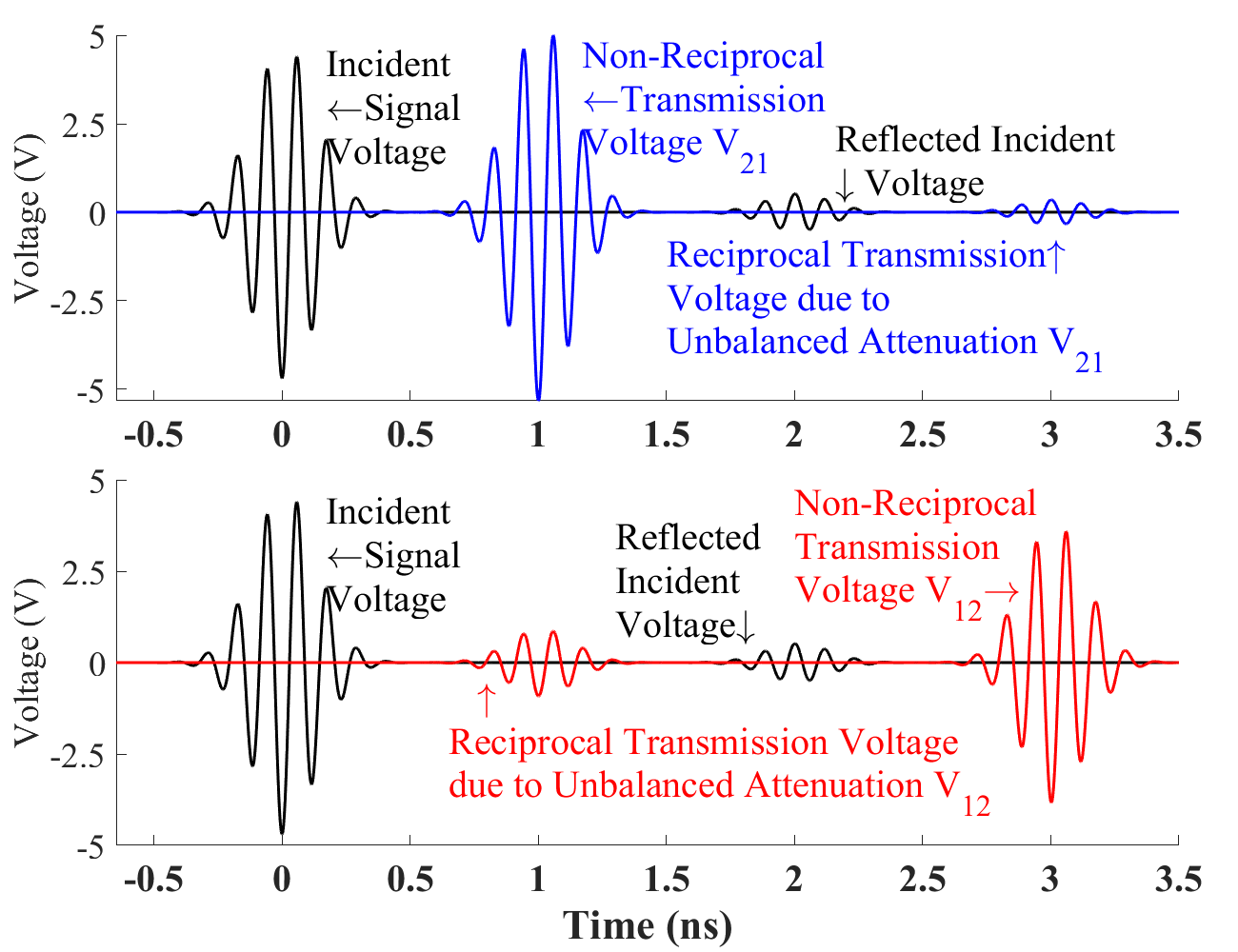}
\caption{CST simulation of pulse propagation through the AB-ring graph in two different directions for the case of attenuation on only one bond (Fig. \ref{schematic_1Att}). The incident signal with center frequency $8.5$ GHz is displayed in black while the two transmitted pulses are displayed in blue for the signal traveling from port 1 to 2 and red for port 2 to 1. The 3:1 time-delay asymmetry is clearly evident, along with a significant difference in pulse amplitudes. The unbalanced attenuation in the ring causes the waves propagating along the two bonds to not have equal magnitudes, hence they do not perfectly interfere at the combiners, causing some degree of reciprocal transmission and a non-zero reflection at the input port.  Compare with Fig. \ref{Sim_Pulse}.}
\label{Sim_Pulse_BottomAttn}
\end{figure}

Figure \ref{Sim_Pulse_BottomAttn} is the time domain analogue of Figure \ref{S_ABRingGraph_BottomAttn}(a). Similar to Figure \ref{Sim_Pulse}, there is both uniform attenuation and $0.35$ Np lumped attenuation in the simulation, but now all the lumped attenuation is located  on one branch. The unbalanced attenuation creates a favored path for the waves to travel along, with a larger amplitude propagating through the branch without the lumped attenuation. When the waves travelling on the un-attenuated upper bond and attenuated lower bond later meet at a tee junction, they do not have the same amplitude. So despite the gyrator adding a $\pi$ phase shift to one side in one direction, the waves don't perfectly interfere and reflect, and a small amount is transmitted. This is the cause of the small blue signal seen at $3$ ns and red signal at $1$ ns in Fig. \ref{Sim_Pulse_BottomAttn}. The reflected incident voltage at $2$ ns is due to the tee junctions having some intrinsic finite reflection, and now due to the unbalanced nature of the graph a small amount of the incident signal will return to escape through its input port, having made a complete return trip through the graph.

Figure \ref{1Atten} shows the experimental results for the asymmetric transmission through a microwave AB-ring graph having un-balanced attenuation $\Gamma_A$ in the bonds, as shown schematically in Fig. \ref{schematic_1Att}.  The experiment includes finite uniform attenuation $\eta$ as well.  Figure \ref{1Atten}(a) shows a schematic of the AB-ring graph with one single attenuator. Figure \ref{1Atten}(b) shows the experimental realization of Figure \ref{1Atten}(a). Figure \ref{1Atten}(c) repeats the asymmetric transmission results obtained from simulations of the quantum system shown in the inset, as presented in the main text.  Figure \ref{1Atten}(d) shows the frequency domain asymmetric transmission results $\braket{P_{21}-P_{12}}$ for both experiment (green diamonds) and simulation (green solid and dashed lines). Figure \ref{1Atten}(e) shows the time domain results $\sigma_V$ for both experiment (red diamonds) and simulation (red solid and dashed lines).

The un-balanced AB-ring-like microwave graph retains all of the asymmetric properties of the fully balanced version, but these features are obscured to some extent by the design of the graph.  Hence even a literal interpretation of the graph shown in Fig. \ref{schematic_ABRing}(a) still shows the same asymmetric transmission properties of the balanced version of the graph.

\textbf{Comments about dephasing models.} Although the work in this paper is strictly classical, it is of interest to discuss models of quantum dephasing, since they originally motivated our work.  The authors of Refs.\cite{Bred21, Mann21} contend that it is advantageous to employ a finite amount of decoherence, implemented through inelastic scattering, to create asymmetric transmission in a model mesoscopic device, suggesting that working in the regime between purely quantum and purely classical physics may be optimal for establishing a net transmission bias that spontaneously pumps charge from one equivalent reservoir to another.  Their model AB ring graph with partial dephasing effectively treats excitations moving in one direction more as waves, while the excitations travelling in the other direction are treated more as particles \cite{ManNR19,Mann21}.  It is an open question whether this approach to creating asymmetric transmission can be implemented in a mesoscopic metal or molecule.  Proposals have been put forward \cite{Mann18,Man2020,Braak20,Bred21,Mann21} that broken time-reversal invariance at the microscopic level can allow one to bypass constraints imposed by the Onsager relations, or restrictions against spontaneous currents \cite{Ohashi96, Mann18}.

  
Other possible experimental settings in the intermediate range between coherent quantum and purely classical behavior (in addition to Ref. \cite{ColdAtom20}) include the following.  
A mesoscopic AB-ring made of a clean metal, or ring-like molecule, with engineered dephasing centers have been proposed to show asymmetric transmission in equilibrium \cite{Let06,Mann21}. 
Another possibility is an AB-ring defined by interfaces supporting topological edge modes, as proposed for layers of buckled silicene \cite{Szafran19}.  Including a defect that creates inter-valley scattering of the electron wavepackets could generate bi-directional asymmetry of transport through the ring.
Finally, it has been proposed that a class of monitored quantum devices can show a non-reciprocal current between reservoirs even in the absence of an applied bias \cite{ferreira2023exact}. 

The results in this paper suggest that simple treatments of dephasing may be effectively equivalent to the PCP approach to describing the effects of loss in classical wave scattering systems.  Although the present work does not address issues in thermodynamics, we note that improper treatment of quantum dephasing have been shown to result in erroneous predictions for the properties of the system, including violations of the second law of thermodynamics \cite{Novo02,Levy14}. 
 Alternative models of quantum dephasing have also been developed, and may be better suited for understanding the properties of mesoscopic systems.  One example of such an approach is the class of collisional models which consider interactions between the quantum system and ancillary environmental qubits, through a series of discrete interactions (collisions)\cite{Barr11,GPIBMQ20}.  Many-body treatments of dephasing have also been pursued \cite{Froml19}.  In general, it would appear that a global quantum treatment of the system and surroundings together is necessary to properly address issues of dephasing \cite{Levy14}.


\clearpage
\newpage

\bibliography{ABRingGraph.bib}

\end{document}